\documentclass[twocolumn,secnumarabic,amssymb, aps,prb,reprint]{revtex4-1}
\usepackage{graphicx}
\usepackage{color}
\usepackage{dcolumn}
\usepackage{bm}
\usepackage{ucs}
\usepackage{physics}

\usepackage{mathtools}
\begin{document}

\title{Two-Impurity Yu-Shiba-Rusinov States in Coupled Quantum Dots}

\author{J. C. Estrada Salda\~{n}a$^{1}$}
\author{A. Vekris$^{1,2}$}
\author{R. \v{Z}itko$^{3,4}$}
\author{G. Steffensen$^{1}$}
\author{P. Krogstrup$^{1,5}$}
\author{J. Paaske$^{1}$}
\author{K. Grove-Rasmussen$^{1}$}
\author{J. Nyg{\aa}rd$^{1}$}

\affiliation{$^{1}$Center for Quantum Devices, Niels Bohr Institute, University of Copenhagen, 2100 Copenhagen, Denmark}
\affiliation{$^{2}$Sino-Danish Center for Education and Research (SDC)
SDC Building, Yanqihu Campus, University of Chinese Academy of Sciences,
380 Huaibeizhuang, Huairou District, 101408 Beijing, China}
\affiliation{$^{3}$Jo\v{z}ef Stefan Institute, Jamova 39, SI-1000 Ljubljana, Slovenia}
\affiliation{$^{4}$Faculty of Mathematics and Physics, University of Ljubljana, Jadranska 19, SI-1000 Ljubljana, Slovenia}
\affiliation{$^{5}$Microsoft Quantum Materials Lab Copenhagen, Niels Bohr Institute, University of Copenhagen, 2100 Copenhagen, Denmark}

\date{\today}

\begin{abstract}
Using double quantum dots as the weak link of a Josephson junction, we realize the superconducting analog of the celebrated two-impurity Kondo model. The device shows a cusped current-voltage characteristic, which can be modelled by an overdamped circuit relating the observed cusp current to the Josephson critical current. The gate dependence of the cusp current and of the subgap spectra are used as complementary ground-state indicators to demonstrate gate-tuned changes of the ground state from an inter-dot singlet to independently screened Yu-Shiba-Rusinov (YSR) singlets. In contrast to the two-impurity Kondo effect in normal-state systems, the crossover between these two singlets is heralded by quantum phase boundaries to nearby doublet YSR phases in which only a single spin is screened.


\end{abstract}

\maketitle

Magnetism relies on the presence of magnetic moments and their mutual exchange interactions. At low temperatures, local moments in metals may be screened by the Kondo effect and magnetism can be disrupted. This competition was first proposed by Mott~\cite{Mott1974} as a mechanism for the vanishing of magnetism at low temperatures in the $f$-electron metal CeAl$_3$, and later explored by Doniach~\cite{Doniach1977} within a simple one-dimensional Kondo-lattice model, from which he established a phase diagram delineating the magnetic phase as a function of the ratio between the Kondo temperature, $T_{K}$, and the inter-impurity exchange. The essence of this competition was subsequently reduced to the vastly simpler two-impurity Kondo model, which exhibits an unstable fixed point separating a ground state (GS) of two local Kondo singlets from an inter-impurity exchange singlet\cite{jones1987study}. This competition remains a central ingredient in the current understanding of many heavy-fermion materials and their quantum critical properties~\cite{georges1996georges,hewson1997kondo,coleman2007heavy,Lohneysen2007,Bulla2008,gull2011continuous}.

\begin{figure} [t!]
\includegraphics[width=1\linewidth]{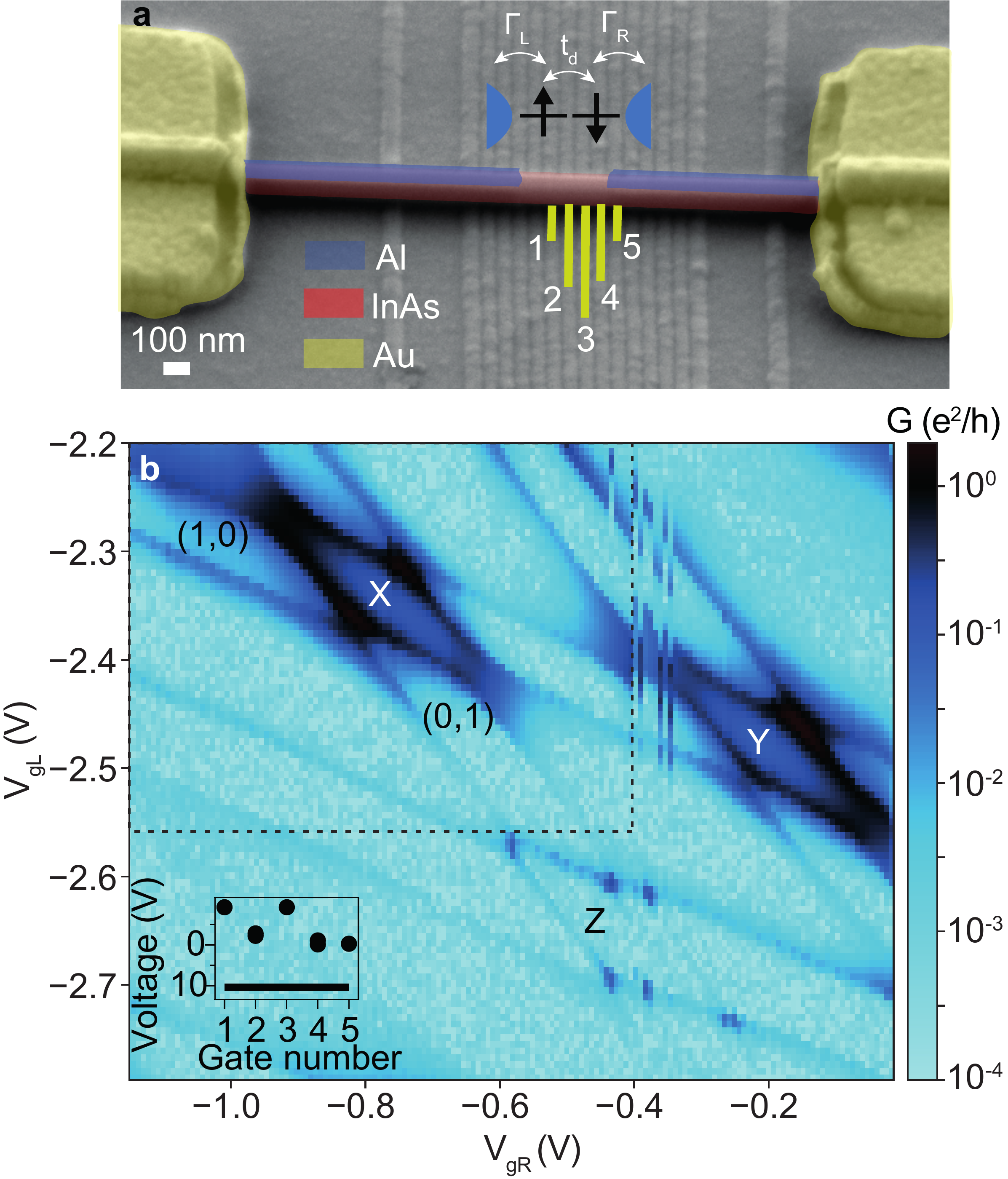}
\caption{(a) Scanning electron micrograph of the device, false-colored to indicate materials. Yellow lines denote local gates used. A sketch of the superconductor-double quantum dot-superconductor (S-DQD-S) system used as a two-impurity YSR model is shown in inset. (b) Zero-bias conductance, $G (V_{\mathrm{gL}},V_\mathrm{{gR}})$, colormap providing an overview of the DQD shell selected for the study (X, inside the dashed line) and surrounding, less stable shells (Y and Z), which are affected by instabilities at $V_\mathrm{{gR}} \approx -0.4$ V. The inset shows gate settings for this measurement. Dots and a line indicate local gate voltages and backgate voltage, respectively. $V_\mathrm{{g1}}=-9.2$ V, $V_\mathrm{{g3}}=-9.2$ V, $V_\mathrm{{g5}}=-0.25$ V and $V_\mathrm{{bg}}=10.4$ V.}
\label{fig1}
\end{figure}

In a superconductor (S), the gap around the Fermi surface terminates the Kondo screening process before its completion, but local magnetic moments may still be screened by forming local singlets with BCS quasiparticles. As demonstrated by Yu, Shiba, and Rusinov (YSR)~\cite{yu1965,shiba1968classical,rusinov1969theory},
a local exchange coupling between a superconductor and a magnetic impurity 
leads to a subgap bound state. In a full quantum description, the corresponding bound state crosses zero energy and the singlet subgap state becomes the new GS at a coupling strength corresponding to $T_K\approx 0.3\Delta$, where $\Delta$ denotes the superconducting gap~\cite{Satori1992,Bauer2007}. 
This quantum phase transition reduces the spin by $\hbar/2$, quenching a spin-1/2 altogether.

Half-filled Coulomb-blockaded quantum dots (QDs) are model magnetic impurities, and can be readily coupled to superconductors in semiconductor nanowires~\cite{Prada2019Nov}. In Josephson junctions (JJs) hosting an impurity, the above GS transition results in a phase change from $\pi$ to 0~\cite{van2006supercurrent,cleuziou2006carbon,grove2007kondo,Jorgensen2007,Maurand2012,delagrange2015manipulating,delagrange2017,saldana2018supercurrent,estrada2018supercurrent,Meden2019Feb}, and constitutes a complementary experimental signature to bias spectroscopy of YSR states crossing zero energy~\cite{deacon2010tunneling,kim2013transport,lee2014spin,jellinggaard2016tuning,lee2017scaling,Gramich2017Nov,grove2017yu,Gramich2017Nov,su2017andreev,EstradaSaldana2020Jul}.
 
In contrast to the normal-state two-impurity Kondo effect, the two-impurity YSR ground-state phase diagram depends strongly on the two different local exchange couplings and includes not only the two different singlets, but also a doublet GS in which only a single spin is screened. In a previous Letter, we used a nanowire device to demonstrate the Josephson effect in a serial double quantum dot (DQD) in the low-coupling regime, in which the inter-dot tunnel coupling, $t_d$, dominated over the dot-lead tunnelling rates, $\Gamma_L, \Gamma_R$~\cite{saldana2018supercurrent}. Here, we investigate in the same device the possible GSs at stronger individual couplings to the leads. As the gate voltages controlling these couplings are tuned, the boundaries of the honeycomb charge stability diagram between GSs of different parity are erased and new two-impurity YSR GSs are accessed. This boundary deletion affects the dispersion in gate voltage of the extracted Josephson critical current, $I_\mathrm{c}$, and of a closely related and directly measurable \textit{cusp current}~\cite{Naaman2001Aug,Rodrigo2004Aug,Randeria2016Apr,Cho2019Jul,Peters2020Jul,Liu2020}, as well as the dispersion of the spectral YSR resonances, which render mutually consistent GS parity information. 

The article is organized in sections as follows. In Section I, we show an overview of the DQD and its parameters. In Section II, we introduce the method used to extract $I_\mathrm{c}$. In Section III, we delineate the theoretical expectations for GS, $I_\mathrm{c}$ and subgap spectra in the S-DQD-S system. In Section IV, we show our main experimental results, making use of the concepts introduced in Section III. Finally, in Section V we present our conclusions and comment on the possible applications of our findings.

\section{Double quantum dot characterization}
 
The measured device (Fig.~\ref{fig1}a) consists of an Al-covered InAs nanowire \cite{krogstrup2015epitaxy} deposited on top of narrow gates, insulated by 20 nm of hafnium oxide from the wire. Al is etched away to form a Josephson junction with a bare segment of InAs nanowire as the weak link. The device has a Si/SiO$_2$ substrate backgate which we operate at $V_{\mathrm{bg}} \sim 10$ V to observe a measurable supercurrent. Gates 1, 3 and 5 are set to negative voltages to define the DQD, while gates 2 (4) are used as the plunger $V_{\mathrm{gL}}$ ($V_{\mathrm{gR}}$) to load electrons into the left (right) QD. Slight changes in the voltage of gate 1 (5) modify $\Gamma_\mathrm{L}$ ($\Gamma_\mathrm{R}$) of the left (right) QD to the left (right) lead. Small changes in $V_{\mathrm{bg}}$ also affect these tunnelling rates. All measurements are done in an Oxford Triton\textsuperscript{\textregistered} dilution refrigerator at $T_{\rm fridge} \approx 20$ mK, using standard lock-in techniques with a lock-in AC excitation of 2 $\mu$V at 116.81 Hz to obtain the differential conductance, $dI/dV_{\mathrm{sd}}$, superposed to a source-drain DC voltage, $V_{\mathrm{sd}}$, while simultaneously recording the current, $I$, with a digital multimeter. The data was corrected for an offset of 3.5 pA from the current amplifier.

Figure \ref{fig1}b shows a zero-bias differential conductance ($G$) colormap which represents a portion of the honeycomb stability diagram of the DQD in the superconducting state. Conductance lines appear at places in which the parity of the system changes, constituting a way to accurately map GS boundaries. We label charge sectors of the DQD shell selected for gate tuning (shell X) by $(N_\mathrm{L},N_\mathrm{R})$, where $N_\mathrm{L}$ and $N_\mathrm{R}$ are the (integer) charges in the highest level of the left and right QDs, respectively. Parity lines separating doublet and singlet regions according to the total number of electrons in both QDs alternate in spacing, consistent with shell filling~\cite{saldana2018supercurrent,jorgensen2008singlet}.
 
From Coulomb-diamond spectroscopy of shell X, we obtain the charging energies of the left and right QDs, $U_\mathrm{L} \approx 1.9$ meV and $U_\mathrm{R} \approx 1.6$ meV. From this spectroscopy, we also find that the level spacing of both QDs is equivalent to their charging energies ($\Delta E_\mathrm{L}\approx 1.6$ meV and $\Delta E_\mathrm{R}\approx 1.8$ meV). $\Delta=0.27$ meV, the parent Al superconducting gap, is found from Coulomb-diamond spectroscopy in deep Coulomb blockade in an opaque regime. The fact that $U_\mathrm{L},U_\mathrm{R}>\Delta$ places the system firmly in the YSR regime~\cite{kirvsanskas2015yu}. From the (0,2)-(1,1) anticrossing in the charge stability diagram, we estimate the inter-dot coupling, $t_\mathrm{d}=0.03-0.05$ meV and the inter-dot charging energy, $U_\mathrm{d}=0.13-0.23$ meV. Measurements of DQD and superconductor parameters are provided in Appendix A (see Figs.~\ref{fig9}, \ref{fig10} and \ref{fig11}).
 
The DQD parameters $\Gamma_\mathrm{L}$ and $\Gamma_\mathrm{R}$ could not be independently measured due to the high critical field ($B_\mathrm{c}=2.1$ T) and high critical temperature ($T_\mathrm{c}=2.2$ K) of the superconducting device. We calculate in Section III for our approximate DQD parameters GS boundaries set at $\Gamma_\mathrm{L} \approx  \Gamma_\mathrm{R} \approx 0.4 < \Delta E_\mathrm{L}, \Delta E_\mathrm{R}$, which provide $\Gamma_\mathrm{L},\Gamma_\mathrm{R}$ sufficiently small to satisfy the single-level approximation~\cite{van2006supercurrent}. Multilevel QD JJs show phase changes at fixed dot charge, irrespective of charge occupation~\cite{van2006supercurrent,delagrange2017}, which is not observed in shells X, Y and Z.

Near charge depletion, QDs are typically strongly confined, but also show the smallest $\Gamma_i/U_i$ ratio, while a sizeable $\Gamma_i/U_i$ ratio for a given temperature is needed to observe a finite Josephson current~\cite{de2010hybrid}. This sets a narrow ($\Delta E_i$, $\Gamma_i$, $U_i$) window which translates into a narrow gate range for single-level DQD JJ operation and $\Gamma_i$ tuning. Outside of this window either multilevel ($\Delta E_i \ll \Gamma_i$), resonant level ($\Gamma_i \gg U_i$), or too weakly-coupled regimes arise ($\Gamma_i \ll U_i$), which is why only a few shells near QD depletion could be studied here.
 
\section{Determination of critical and cusp currents}
 
\begin{figure} [b!]
\includegraphics[width=1\linewidth]{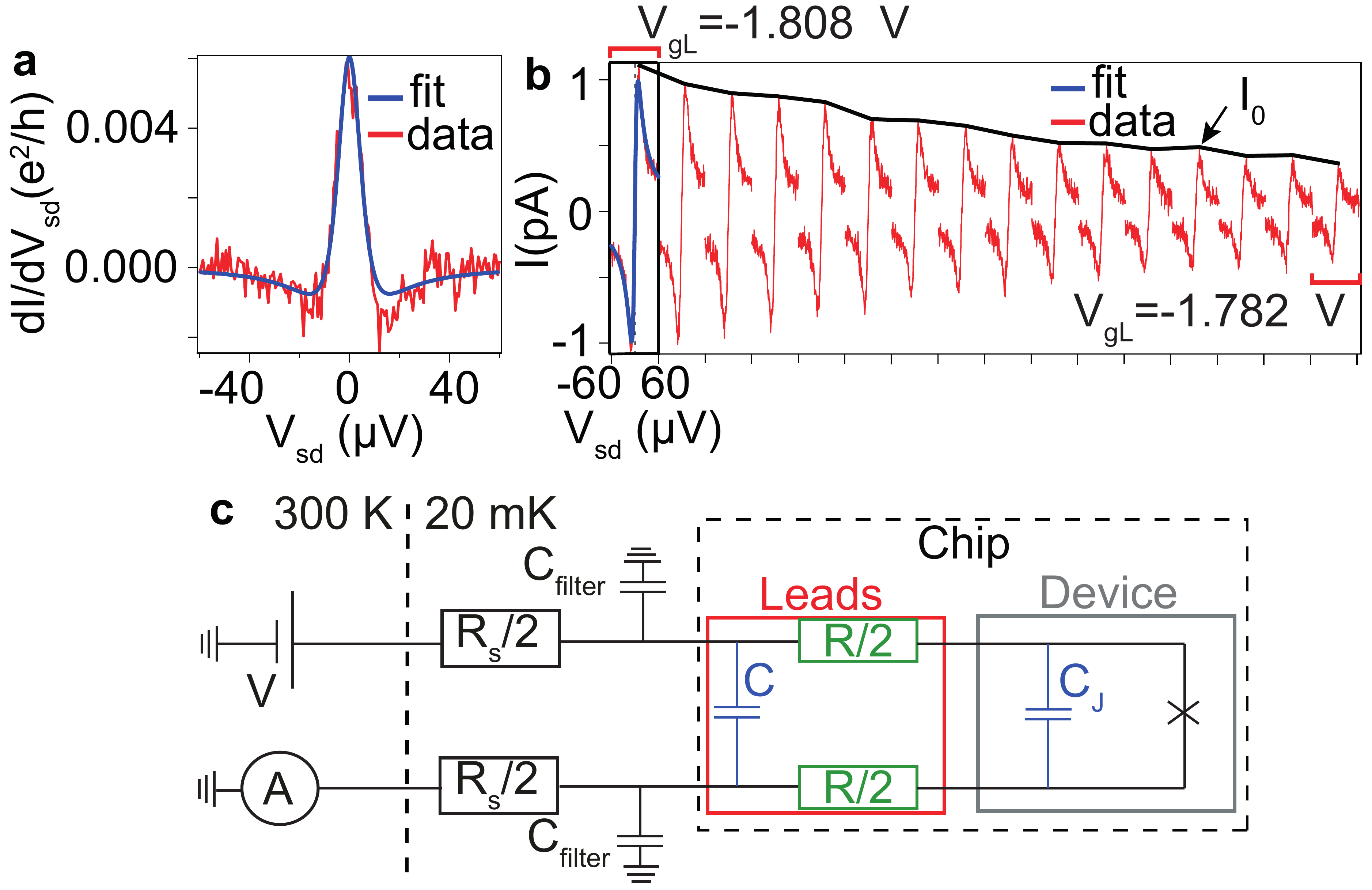}
\caption{(a) $dI/dV_{\mathrm{sd}}-V_{\mathrm{sd}}$ trace recorded at the same gate voltage as the leftmost trace in (b). (b) $I-V_{\mathrm{sd}}$ traces taken with the gates swept along the red solid line in Fig.~\ref{fig8}c. Initial and final gate voltages are indicated. The traces have been shifted horizontally by 120 $\mu$V with respect to each other, and are measured from -60 to 60 $\mu$V each. Cusp current, $I_\mathrm{0}$, at positive $V_{\mathrm{sd}}$ is plotted as a black trace to highlight its gate dependence. (c) Electrical circuit model. Details are given in the text and in Appendix B.}
\label{fig2}
\end{figure}

Next, we focus on the low-bias transport characteristic of the device in the superconducting state.

Initially intended for two-terminal bias spectroscopy of YSR excitations in a S-DQD-S system, for sufficiently strong $t_\mathrm{d}$ the device exhibits in $dI/dV_{\mathrm{sd}}-V_{\mathrm{sd}}$ traces a bias-symmetric, 20 $\mu$V-wide zero-bias peak flanked by negative differential conductance (NDC) dips, exemplified in Fig.~\ref{fig2}a. As shown in Fig.~\ref{fig2}b, the corresponding $I-V_{\mathrm{sd}}$ traces have a finite slope between sharp cusps which occur at a current $I_\mathrm{0}$. The positive $V_{\mathrm{sd}}$ value at which the cusps occur, $V_0=9.1 \pm 0.6$ $\mu$V, is gate-independent over 3 orders of magnitude of $G$ and over cusp currents, $I_\mathrm{0}$, ranging from 0.3 pA to 50 pA (see Fig.~\ref{fig12} in Appendix B). 
Additionally, $V_\mathrm{0}$ matches the $V_{\mathrm{sd}}$ value of NDC dips, and $I_\mathrm{0}$ shows a gate dependence similar to $G$ (for example, compare Figs.~\ref{fig14}c,d in Appendix C), indicating common origins for both phenomena.

We relate this narrow, bias-symmetric, finite-sloped $I-V_{\mathrm{sd}}$ characteristic, which is cusped at $I_\mathrm{0}$ and locked at $V_\mathrm{0}$, to the supercurrent of the circuit-damped JJ formed by the S-leads and the DQD weak-link, schematized in Fig.~\ref{fig2}c. This minimal model assumes the presence of a resistance $R$ at the level of the leads (indicated in green) which was not designed and which was not independently measured, and which provides Johnson-Nyquist white noise at a temperature $T$ which in principle can be larger than the temperature in the dilution refrigerator~\cite{Pekola2010Jul,Naaman2001Aug}. From its fitted value and position in the model, we speculate that $R$ comes from the contact resistance between the nanowire and the Ti/Au leads.

In Fig.~\ref{fig2}c, the designed elements of the circuit at 20 mK are indicated in black. The JJ is assumed to have a current-phase relationship $I_\mathrm{c} \sin \phi$, where the critical current, $I_\mathrm{c}$, is unknown. This relationship is valid away from phase transitions~\cite{saldana2018supercurrent}.
Standard filtering and shielding elements include: 1) Two-stage low-pass RC filters with 30 kHz cut-off frequency, of total resistance $R_\mathrm{s}/2$ per line, where $R_\mathrm{s}=8.24$ k$\Omega$, and capacitance per stage $C_{\rm filter}=2.7$ nF.
2) 7-stage low-pass $\pi$ filters with 200 MHz cut-off frequency, positioned at 20 mK in series with the RC filters (not shown).
3) Lines to the sample are additionally attenuated at high frequency by Eccosorb\textsuperscript{\textregistered} and encased by copper tape. 
4) Radiation shielding is used at several stages in the dilution refrigerator, with the deepest encasing being at 20 mK at the sample level.

Non-designed on-chip elements of known origin in the experiment are indicated in blue. Using a parallel-plate capacitor model, and from the device geometry and materials, we evaluate the junction capacitance at $C_\mathrm{J} \approx 3$ aF, and the capacitance of the large-area bonding pads through the backgate at $C \approx 9$ pF. The latter estimation assumes that the resistance R is somewhere in the leads between the bonding pads and the device. These capacitances place the JJ in the overdamped regime, but knowledge of their values is not needed to extract $I_\mathrm{c}$.

In the overdamped and voltage-biased regime this circuit gives rise to a three-parameter ($I_\mathrm{c}$, $R$, $T$) voltage-current relation, as demonstrated by Ivanchenko and Zil'berman~\cite{Anchenko1969}. When $T> \hbar I_\mathrm{c}/2ek_B$, corresponding to the limit of large noise amplitude~\cite{Naaman2001Aug}, $V_\mathrm{0}$ is $I_\mathrm{c}$-independent as in the experiment and its value $V_\mathrm{0}=2eRk_BT/ \hbar=10$ $\mu$V is close to the measurement. Additionally, $I_\mathrm{c}$ can be rescaled to $I_\mathrm{0}$ and $G$, as shown in Fig.~\ref{fig13} in Appendix B, explaining their connection in the experiment. 

We fit $dI/dV_{\mathrm{sd}}-V_{\mathrm{sd}}$ and $I-V_{\mathrm{sd}}$ curves to obtain $I_\mathrm{c}$. Since the device is measured in a two-terminal configuration, $R_\mathrm{s}=8.24$ k$\Omega$ is subtracted from the raw bias voltage, $V_{\mathrm{sd}}=V-IR_\mathrm{s}$, and raw differential conductance data, $dI/dV_{\mathrm{sd}}$=$(dI/dV)/(1-R_\mathrm{s}dI/dV$), before inputting them into the formula. Despite the circuit assumptions and the crudity of the model employed, the fits to the data are good for $I_\mathrm{0}$ between 0.2 pA and 50 pA, and $I_\mathrm{c}$ ranging from 0.03 nA to 0.8 nA. Examples of the fit are shown as blue curves in Figs.~\ref{fig2}a,b.


The parameters $R=3$ k$\Omega$ and $T=80$ mK are extracted through the fitting procedure by initially keeping the parameters $I_\mathrm{c}$, $R$ and $T$ free. As long as $RT=240$ k$\Omega \cdot$mK, $R$ and $T$ can adopt any value and still produce a good fit to the data. A rescaling by a dimensionless number $\gamma$ to $R/\gamma$ and $T\gamma$ globally will merely change $I_\mathrm{c}$ to $\sqrt{\gamma}I_\mathrm{c}$ without any qualitative change in its gate dependence. This is crucial for a robust interpretation of the latter, but it prevents absolute quantitative $I_\mathrm{c}$ assessments. Nevertheless, $T$ is chosen from an upper-bound estimate of the electron temperature from the thermal broadening of a Coulomb peak, which also produces a reasonable $I_\mathrm{c}$ value, given the maximum imposed by $\Delta$ and the DQD couplings. The fitted value of $R$, in turn, is compatible with previous measurements of contact resistance in highly transparent nanowire junctions~\cite{vanWeperen2013Feb}. Once the noise $RT$ was set, $I_\mathrm{c}$ was kept as the only free parameter in the fit to $V_{\mathrm{sd}}-I$ curves at other gate voltages. 
Extraction of an unambiguous $I_\mathrm{c}$ value, impeded here by lack of measurement of $R$~\cite{Naaman2001Aug}, should benefit from a fully designed circuit~\cite{steinbach2001direct}.
A derivation of the model and additional details are shown in Appendix B.

 \begin{figure} [b!]
\includegraphics[width=1\linewidth]{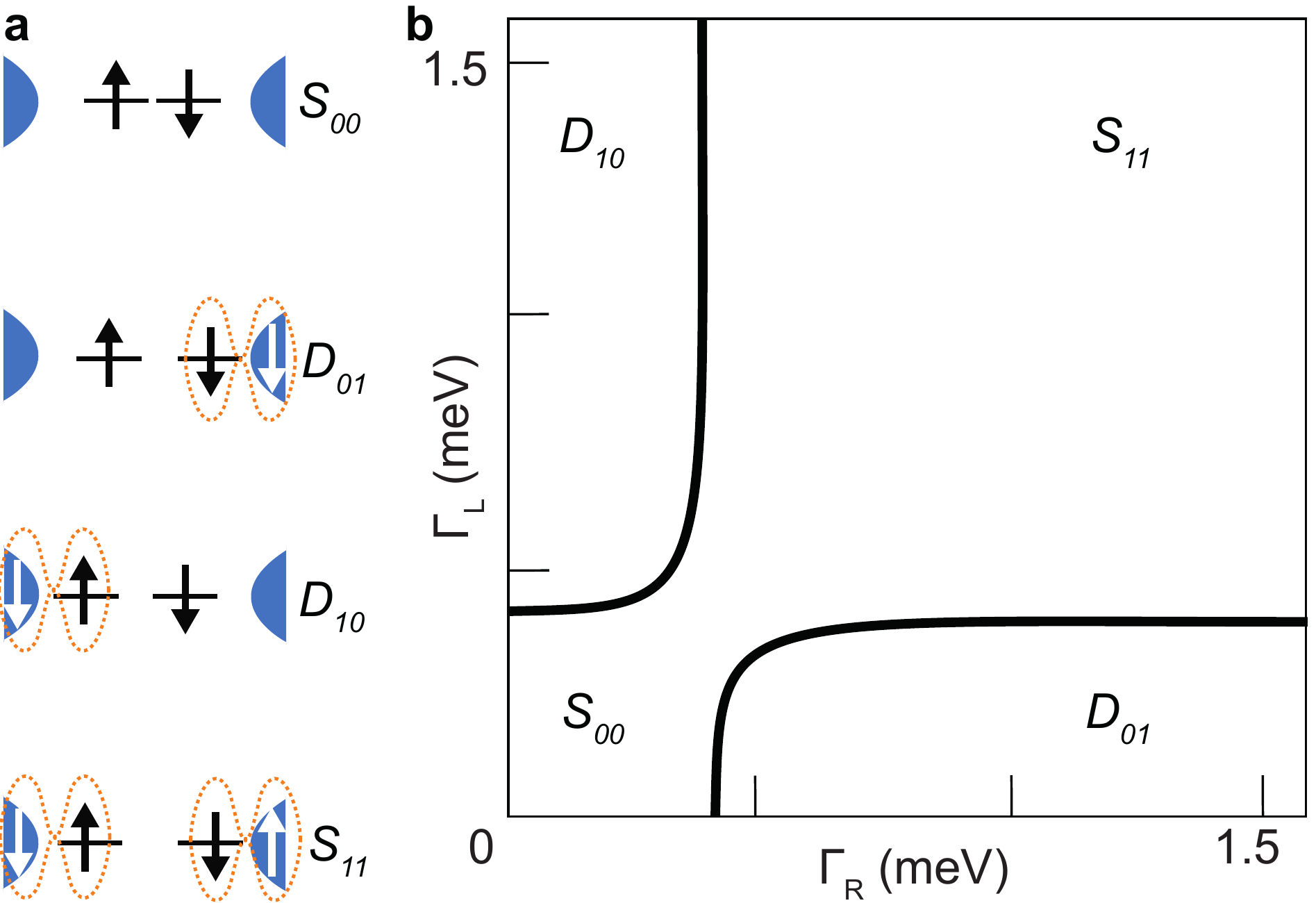}
\caption{(a) Schematized two-impurity YSR interaction (orange dotted line) of electron spins (black arrows) and quasiparticles (white arrows) from superconducting leads for different tunnelling rates $\Gamma_\mathrm{L}, \Gamma_\mathrm{R}$. Horizontal lines indicate double quantum dot (DQD) levels. Each level contains one electron. (b) NRG calculation of the phase diagram of the two-impurity YSR S-DQD-S system for gates $n_\mathrm{L}=n_\mathrm{R}=1$ corresponding to single occupancy of both dots.
}
\label{fig3}
\end{figure}

\begin{figure} [b!]
\includegraphics[width=1\linewidth]{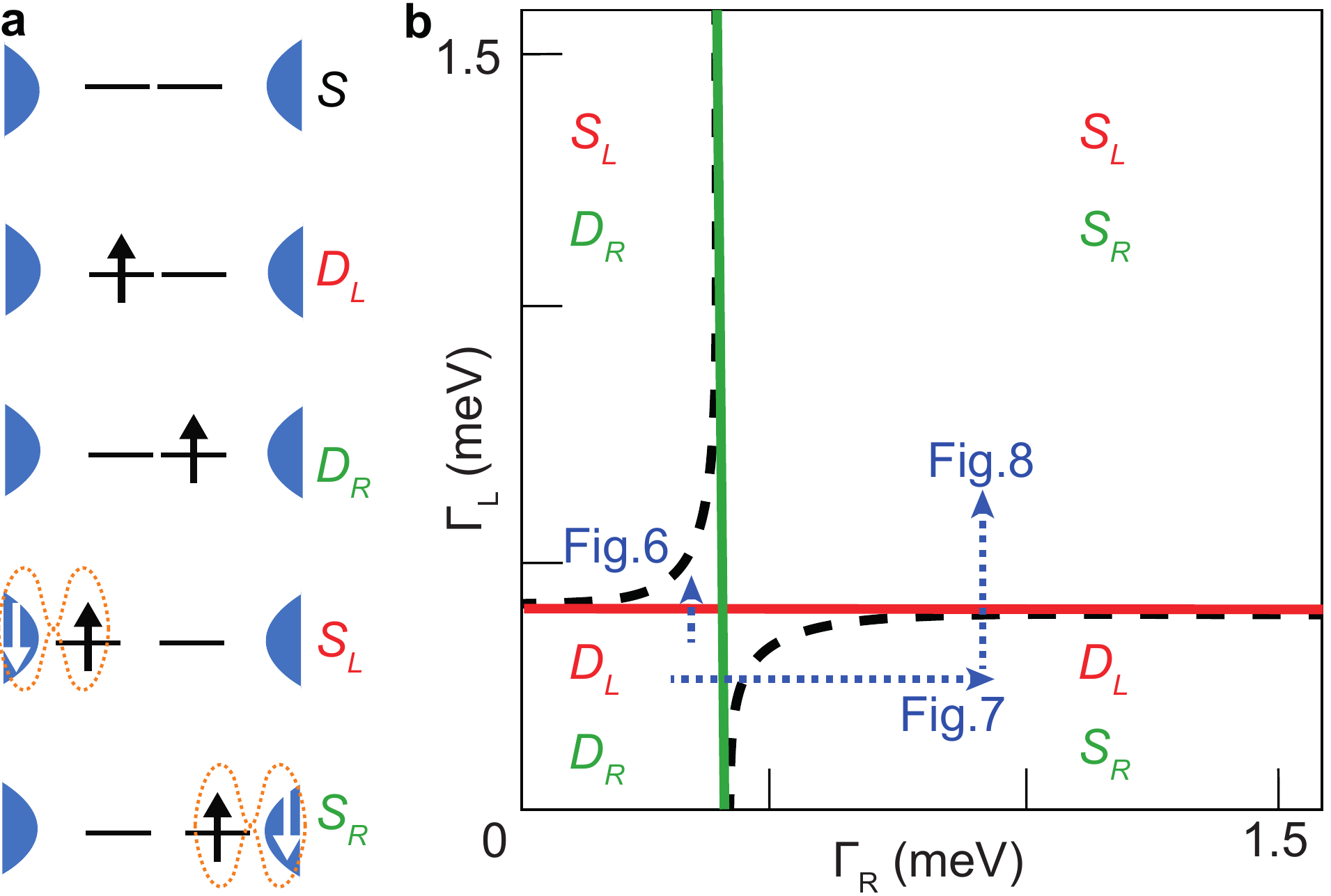}
\caption{(a) Schematized YSR interaction of electron spins and quasiparticles from superconducting leads for different $\Gamma_\mathrm{L}, \Gamma_\mathrm{R}$ and level occupations. The bottom four cases correspond to the one-impurity YSR system. (b) NRG calculation of the phase diagram of the one-impurity YSR S-DQD-S system for single occupancy of either dot. GS boundaries for $n_\mathrm{L}=1$, $n_\mathrm{R}=0$ are denoted by a red solid line, while those for $n_\mathrm{L}=0$, $n_\mathrm{R}=1$ are denoted by a green solid line. Dotted blue arrows indicate qualitatively the gate-tuned $\Gamma_\mathrm{L}, \Gamma_\mathrm{R}$ paths in the sequences of experimental charge stability diagrams in Figs.~\ref{fig6}, \ref{fig7} and \ref{fig8}.
}
\label{fig4}
\end{figure}

\begin{figure} [t!]
\includegraphics[width=1\linewidth]{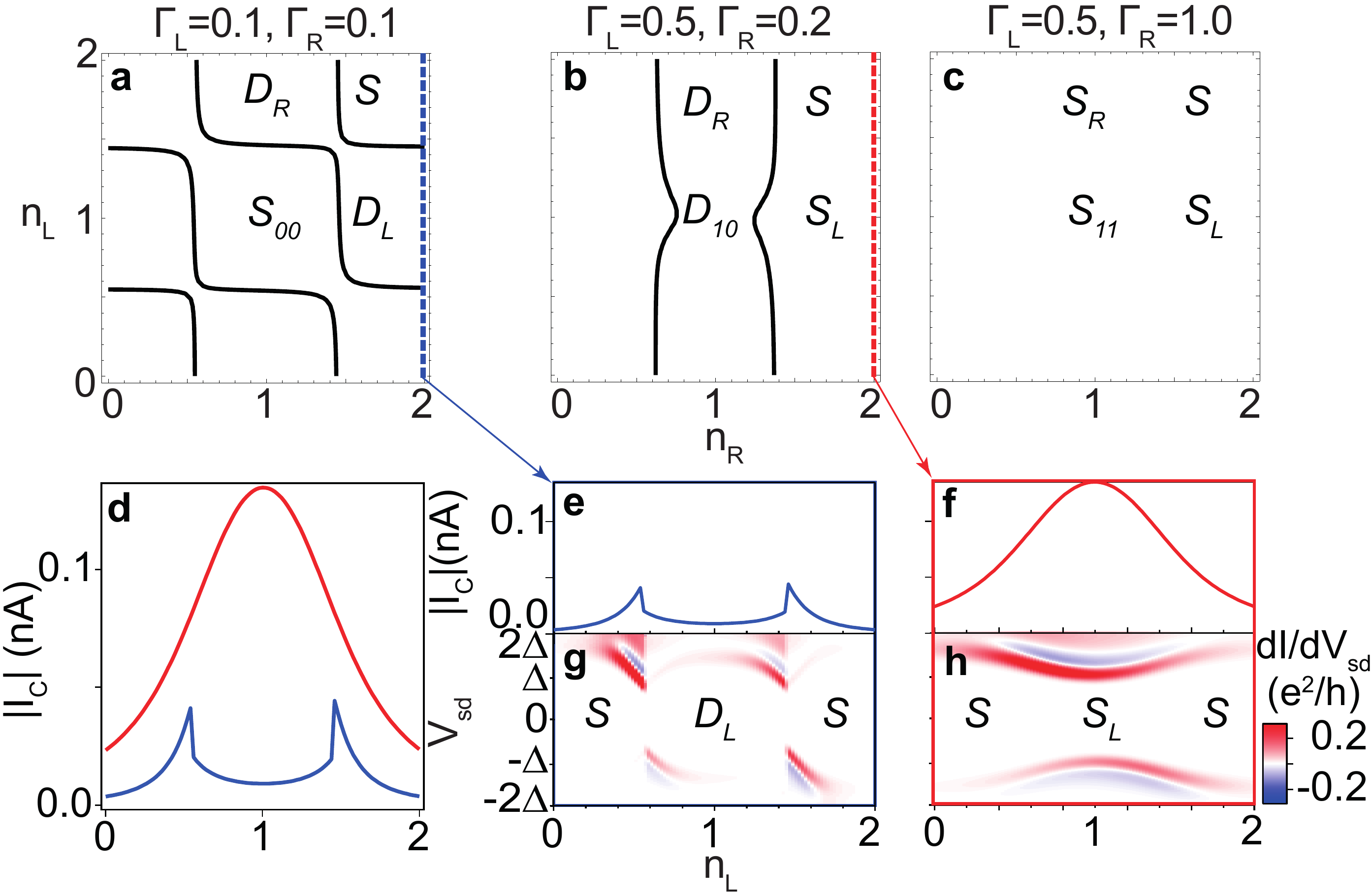}
\caption{NRG calculations. (a-c) Charge stability diagram with doublet-singlet GS boundaries indicated as solid black lines. (d) $\abs{I_\mathrm{c}}$ along respectively colored dashed lines in (a,b). (e,f) Comparison of the gate dependence of $\abs{I_\mathrm{c}}$ to (g,h) colormaps of subgap $dI/dV_{\mathrm{sd}}-V_{\mathrm{sd}}$ along the same gate trajectory. The colormaps are calculated by the Bardeen approach (see Appendix E).}
\label{fig5}
\end{figure}

\section{Theoretical insights}

We introduce below the phase diagram of the S-DQD-S system, which distills the main consequence of the YSR spin-screening mechanism: changes in tunnelling rates $\Gamma_L$ and $\Gamma_R$ drive GS parity transitions. We also provide calculations of the parity stability diagram, of $I_c$, and of the subgap $dI/dV_{\mathrm{sd}}$ YSR spectra, of importance to undertake the experimental results in the next section.

In our GS naming convention, the parity is specified by $D$ for doublet and $S$ for singlet. Indexes depend on QD occupation, as follows. If the QD levels contain each one electron, we use two indexes; the first (second) denotes if the left (right) spin is screened (1) or not (0) (Fig.~\ref{fig3}a). If only one QD contains a single electron and the other is empty/full, we use a single index, L or R, to indicate if the left or right QD level is occupied (Fig.~\ref{fig4}a). Finally, if both QD levels are empty/full, we use no index. For example, in Fig.~\ref{fig3}a $D_\mathrm{01}$ denotes screening of the right-QD spin for the (1,1) charge state.



In Fig.~\ref{fig3}b, we show the two-impurity phase diagram of the S-DQD-S system vs.~$\Gamma_\mathrm{L}$, $\Gamma_\mathrm{R}$ with dot occupations fixed at $n_\mathrm{L}=n_\mathrm{R}=1$.
This diagram was calculated by the Numerical Renormalization Group (NRG) technique~\cite{wilson1975,lee2014spin} for $t_\mathrm{d}=0.1$ meV, charging energies of the QDs $U_\mathrm{L}=U_\mathrm{R}=2$ meV, and $\Delta=0.25$ meV, parameters which are used in all NRG calculations shown in this work. For simplicity, $U_\mathrm{d}$ was kept at zero. Black lines denote GS parity boundaries between $S_{\mathrm{00}}$, $S_{\mathrm{11}}$ and $D_{\mathrm{10}}$, $D_{\mathrm{01}}$. These boundaries avoid each other due to $t_\mathrm{d}$.
An NRG calculation of the one-impurity phase diagram is in turn shown in Fig.~\ref{fig4}b. The one-impurity case corresponds to $n_\mathrm{L}=1$, $n_\mathrm{R}=0,2$ (red) and to $n_\mathrm{L}=0,2$, $n_\mathrm{R}=1$ (green). In contrast to two-impurity GS parity boundaries (black dashed lines overlaid from the calculation in Fig.~\ref{fig3}b), one-impurity parity boundaries are straight lines.

\begin{figure*} [t!]
\includegraphics[width=1\linewidth]{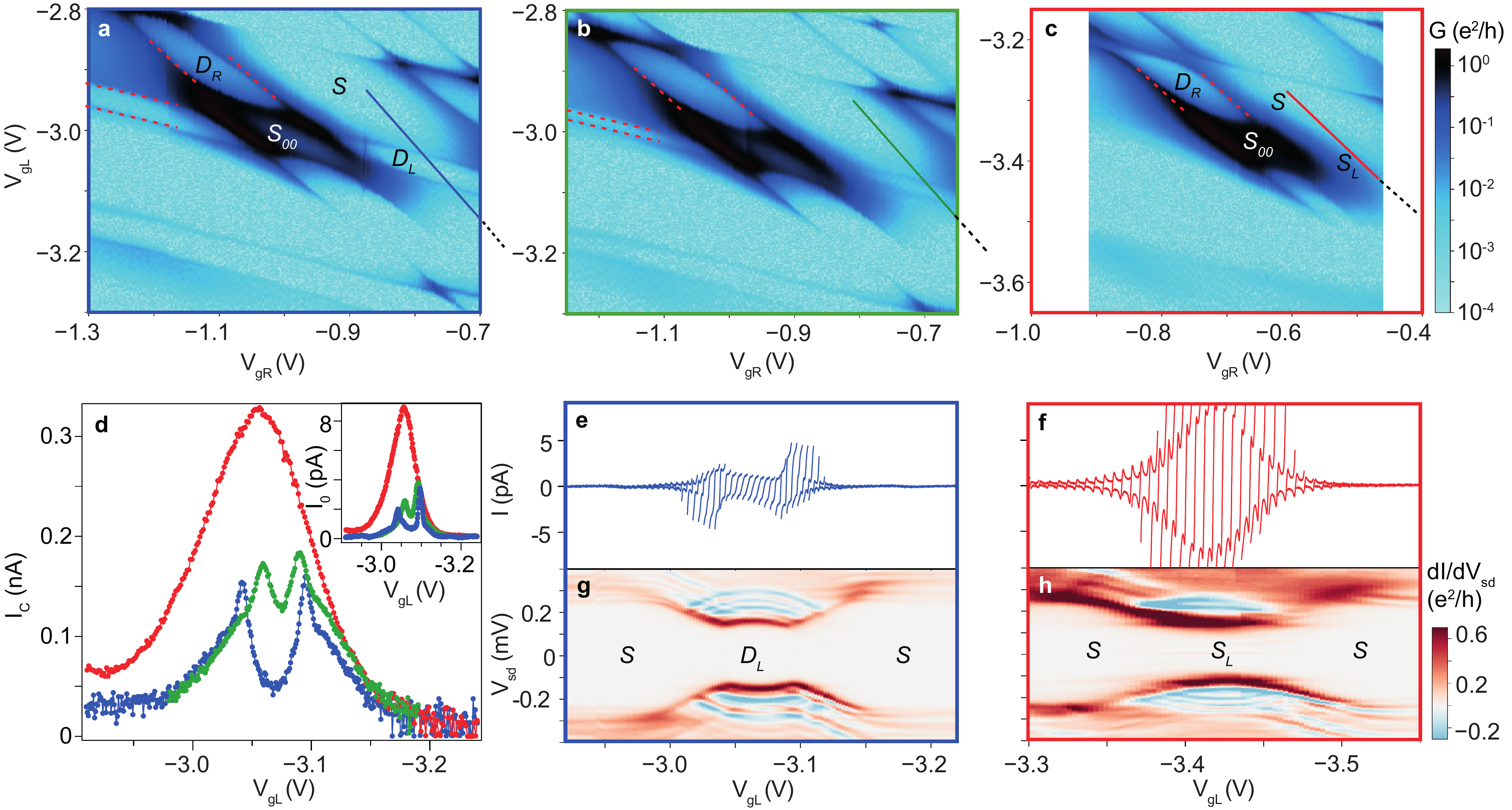}
\caption{(a-c) $G (V_{\mathrm{gL}}, V_{\mathrm{gR}})$ colormaps taken with $V_{\mathrm{g1}}$ set to (b) -9.2 V, (c) -9 V, (d) -8.95 V. In (a,b), other gates are set to $V_{\mathrm{g3}}= -9.3$ V, $V_{\mathrm{g5}}= -0.25$ V and $V_{\mathrm{bg}}= 10.4$ V. In (c), other gates are set to $V_{\mathrm{g3}}= -9.0$ V, $V_{\mathrm{g5}}= -0.36$ V and $V_{\mathrm{bg}}= 11$ V. Changes in $V_{\mathrm{g1}}$ by 0.2 V require compensation in $V_{\mathrm{gR}}$ by $\approx 0.03$ V and in $V_{\mathrm{gL}}$ by $\approx 0.04$ V to keep the charge stability diagram in frame. (d) $I_\mathrm{c}$ and $I_\mathrm{0}$ (in inset) vs.~$V_{\mathrm{gL}}$, with $V_{\mathrm{gL}}$ and $V_{\mathrm{gR}}$ swept along the solid lines of corresponding color in (a-c). A black dotted line after the solid lines in (a-c) indicates that the linecuts extend beyond the gate range of the colormap. The traces are horizontally shifted to match the same gate range. (e,f) Comparison of the gate dependence of $I-V_{\mathrm{sd}}$ traces from which the $I_\mathrm{c}$ and $I_\mathrm{0}$ data in the blue and red curves in (d) are extracted to (g,h) colormaps of subgap $dI/dV_{\mathrm{sd}}-V_{\mathrm{sd}}$ measured along these gate trajectories. To avoid crowding, in (e,f) only every fifth trace is plotted, and the traces are horizontally shifted by 65 $\mu$V with respect to each other. $V_{\mathrm{sd}}$ is swept between $\pm 60$ $\mu$V in each trace.}
\label{fig6}
\end{figure*}

GS parity boundaries can also be plotted as function of $n_\mathrm{L}$, $n_\mathrm{R}$ for fixed $\Gamma_\mathrm{L}$, $\Gamma_\mathrm{R}$, which encourages qualitative comparison to experimental charge stability diagrams such as the one in Fig.~\ref{fig1}b. Three examples calculated by NRG are shown in Fig.~\ref{fig5}. Similarly to the charge stability diagram of shell X in Fig.~\ref{fig1}b, the GS diagram of Fig.~\ref{fig5}a exhibits all possible parity boundaries; its $\Gamma_\mathrm{L}$, $\Gamma_\mathrm{R}$ coordinates situate it at the bottom-left corner of the two and one-impurity phase diagrams of Figs.~\ref{fig3}b and \ref{fig4}b. In Fig.~\ref{fig5}b, at stronger $\Gamma_\mathrm{L}$, the left spin is screened and the GS changes in the (1,0) and (1,2) charge sectors from $D_{\mathrm{L}}$ to $S_{\mathrm{L}}$, and in the (1,1) sector from $S_{\mathrm{00}}$ to $D_{\mathrm{10}}$, deleting the horizontal parity boundaries which previously existed in Fig.~\ref{fig5}a between $S_{\mathrm{00}}$ and $D_{\mathrm{R}}$, and between $D_{\mathrm{L}}$ and $S$. In Fig.~\ref{fig5}c, at stronger $\Gamma_\mathrm{R}$ than in Fig.~\ref{fig5}b, the remaining doublet sector is screened away into a singlet, resulting in the disappearance of vertical parity boundaries.

Modifications of the GS diagram result in changes of the gate dependence of $I_\mathrm{c}$ and subgap YSR spectra which are also experimentally resolvable. An NRG calculation of $\abs{I_\mathrm{c}}$ is shown in Fig.~\ref{fig5}d for an $n_\mathrm{L}$ trajectory which crosses two GS parity boundaries in the case of the blue trace, and no GS parity boundaries in the case of the red one. The blue trace shows asymmetric peaks located at the two $D_{\mathrm{L}}-S$ GS parity changes, whereas the red one evolves into a smooth and broad peak centered at $n_\mathrm{L}=1$. These are the well-known signatures of a $\pi$ and a $0$ QD JJ, respectively. In Figs.~\ref{fig5}e-h, we compare the gate dependence of these traces (Figs.~\ref{fig5}e,f) to a calculation of subgap $dI/dV_{\mathrm{sd}}$ (Figs.~\ref{fig5}g,h), which is detailed in Appendix E. In Figs.~\ref{fig5}g,h, the calculation shows $dI/dV_{\mathrm{sd}}$ colormaps in which the innermost red lines appear at $eV_{\mathrm{sd}}=\pm (\Delta+E_\mathrm{L})$, corresponding to quasiparticle tunnelling from the gap edge of the right superconductor via the spinless right QD, to the YSR state at energy $E_\mathrm{L}$ in the spinful left QD, with tunnelling rate $|t_\mathrm{d}|^2$. Since BCS peaks probe sub-gap bound states, these peaks are followed by dips of NDC. This transport channel assumes a quasiparticle relaxation rate in the left superconductor, which is faster than the inter-dot tunneling rate~\cite{Ruby2015Aug}.
In Fig.~\ref{fig5}g, the YSR $dI/dV_{\mathrm{sd}}$ peaks form the well-known split-loop which kinks downwards at parity changes coinciding with peaks in $\abs{I_\mathrm{c}}$ in Fig.~\ref{fig5}e. In the absence of parity changes as in Fig.~\ref{fig5}h, the YSR peaks evolve smoothly with minimum splitting at $n_\mathrm{L}=1$, which coincides with a broad peak in $\abs{I_\mathrm{c}}$ in Fig.~\ref{fig5}f.

Since the high magnetic field and/or temperature needed to drive the device into the normal state impede independent measurement of $\Gamma_\mathrm{L}$ and $\Gamma_\mathrm{R}$, these calculations serve only as a qualitative guide for the gate-tuned behavior of experimental charge stability diagrams and of the gate dependence of $I_\mathrm{c}$, $I_\mathrm{0}$ and YSR excitations in the section below.

\section{Results}

\begin{figure*} [t!]
\includegraphics[width=0.85\linewidth]{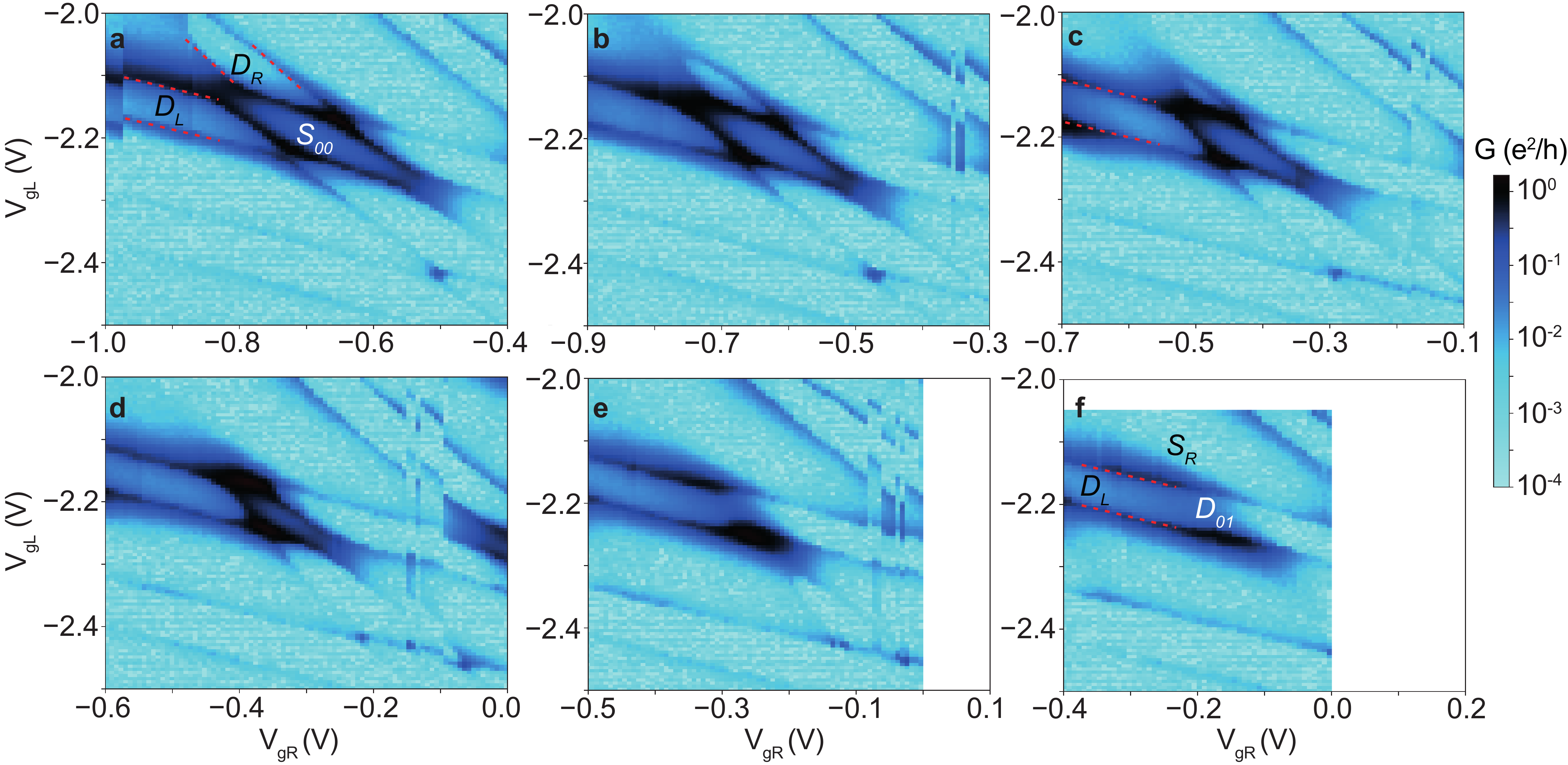}
\caption{Electron-hole asymmetric YSR screening. (a-f) $G (V_{\mathrm{gL}}, V_{\mathrm{gR}})$ colormaps with $V_{\mathrm{g5}}$ set at (a) -0.5 V, (b) -0.75 V, (c) -1.25 V, (d) -1.5 V, (e) -1.75 V, and (f) -2 V. Other gates are set to $V_{\mathrm{g1}}=-9.2$ V, $V_{\mathrm{g3}}=-9.2$ V, $V_{\mathrm{bg}}=10$ V. Changes in $V_{\mathrm{g5}}$ by -0.25 V require compensation in $V_{\mathrm{gR}}$ by $\approx 0.1$ V, an in $V_{\mathrm{gL}}$ by $\approx 0.008$ V to keep the charge stability diagram in frame. The sequence of parity stability diagrams serves as a prelude for Fig.~\ref{fig8}.} 
\label{fig7}
\end{figure*}

\begin{figure*} [t!]
\includegraphics[width=1\linewidth]{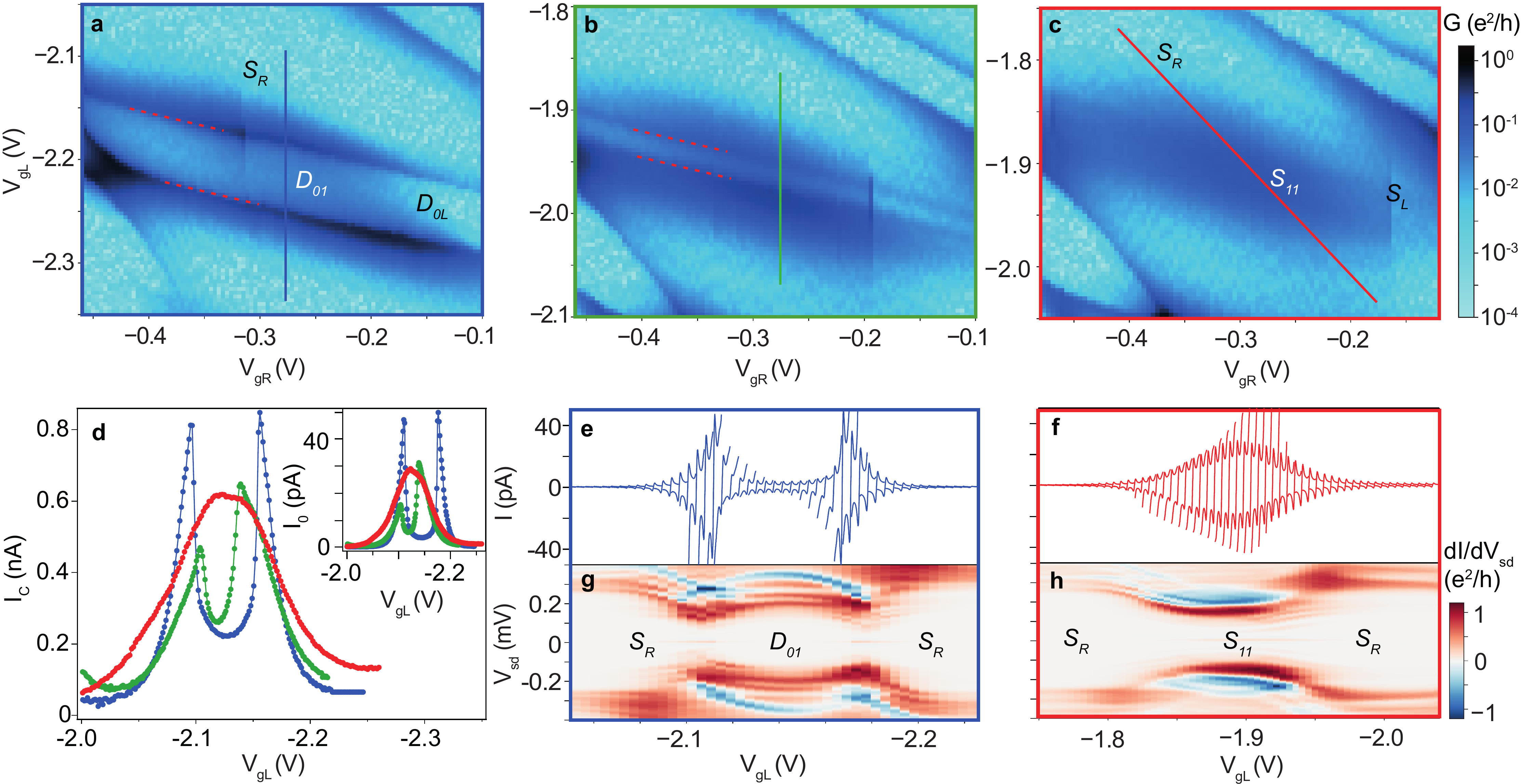}
\caption{(a-c) $G (V_{\mathrm{gL}}, V_{\mathrm{gR}})$ colormaps taken with $V_{\mathrm{bg}}$ set to (a) 10 V,  (b) 9.74 V, and (c) 9.6 V. Other gates are set to $V_{\mathrm{g1}}=-9.2$ V, $V_{\mathrm{g3}}=-9.2$ V, $V_{\mathrm{g5}}=-2$ V. Changes in $V_{\mathrm{bg}}$ by -0.4 V require compensation in $V_{\mathrm{gL}}$ by $\approx 0.3$ V to keep the charge stability diagram in frame.
(d) $I_\mathrm{c}$ and $I_\mathrm{0}$ (in inset) vs.~$V_{\mathrm{gL}}$, with $V_{\mathrm{gL}}$ and $V_{\mathrm{gR}}$ swept along the solid lines of corresponding color in (a-c). The traces are horizontally shifted to match the same gate range. (e,f) Comparison of the gate dependence of $I-V_{\mathrm{sd}}$ traces from which the $I_\mathrm{c}$ and $I_\mathrm{0}$ data in the blue and red curves in (d) are extracted to (g,h) colormaps of subgap $dI/dV_{\mathrm{sd}}-V_{\mathrm{sd}}$ measured along these gate trajectories. To avoid crowding, in (e,f) only every third trace is plotted, and the traces are horizontally shifted by 65 $\mu$V with respect to each other. $V_{\mathrm{sd}}$ is swept between $\pm 60$ $\mu$V in each trace. The sequence of gate-tuned parity stability diagrams, which started in Fig.~\ref{fig7}a in a honeycomb configuration, culminates here in a configuration devoid of parity lines.}
\label{fig8}
\end{figure*}

 By following the deletion of parity transition lines in the charge stability diagram for DQD-shell X, while knowing the departure ground states of each charge configuration $(N_\mathrm{L},N_\mathrm{R})$ of the original diagram at weak-coupling, it is possible to track GS changes of the system at stronger coupling to the leads, to which we access by changing gate voltages other than $V_{\mathrm{gL}}$ and $V_{\mathrm{gR}}$. Before describing the data, we discuss three points not included in the NRG model above:
 
 \begin{enumerate}
     \item While the model is based on a single DQD shell, the DQD device has a staircase of shells. GS changes demonstrated here for shell X are not necessarily concurrent in shells Y and Z, as it is typically the case in nanowire-defined QD devices~\cite{jellinggaard2016tuning,estrada2018supercurrent}.
     \item Parity transition lines in the experimental charge stability diagram of shell X have an acute angle with respect to each other due to trivial $V_{\mathrm{gL}}$, $V_{\mathrm{gR}}$ cross-talk~\cite{vanderWiel2002Dec}, instead of the nearly perpendicular angle which they exhibit in the NRG calculation. While knowledge of the exact angle is not relevant, it is important to keep track of the slope of the parity transition lines of both QDs during the experiment, to avoid confusing gate-controlled parity changes with the formation of a single QD or with the introduction of a third QD, both of which should have parity transition lines of different slope. We therefore keep the same total gate variations $\Delta V_{\mathrm{gL}}$ and $\Delta V_{\mathrm{gR}}$ and aspect ratio in the stability diagrams of Figs.~\ref{fig6}, \ref{fig7} and \ref{fig8}, and use red dashed lines to mark the slope of the left and right QD parity lines. These lines have the same slope in all plots, showing that the device stays as a DQD through the whole gating procedure.
     \item As discussed below, electron-hole asymmetric GS parity changes in shell X are observed for YSR screening of the right-QD spin, which can be modelled by a dependence of $\Gamma_\mathrm{R}$ on $V_{\mathrm{gL}}$ (see Appendix F). This cross-coupling is not unreasonable in view of the slight horizontal rightwards shift of the bottom gates in Fig.~\ref{fig1}a with respect to the nanowire channel~\cite{estrada2018supercurrent}.
 \end{enumerate}

In Fig.~\ref{fig6}a, we show a zero-bias conductance colormap which represents the charge stability diagram of DQD shell X at slightly different gate settings than in the overview in Fig.~\ref{fig1}b. In these settings, the DQD has been brought to the verge of a GS transition due to YSR screening of the left-QD spin. 
We tune the leftmost gate of the device, $V_{\mathrm{g1}}$, to gradually merge (Fig.~\ref{fig6}b) and then, in combination with slight changes in $V_{\mathrm{g3}}$, $V_{\mathrm{g5}}$ and $V_{\mathrm{bg}}$, to erase (Fig.~\ref{fig6}c) the parity transition lines of the stability diagram corresponding to the loading of a spin-1/2 in the left QD, consistent with an increase in $\Gamma_\mathrm{L}$. The end result is that the singlet-doublet-singlet GS sequence $S-D_{\mathrm{L}}-S$ along the solid blue line in Fig.~\ref{fig6}a changes into an all-singlet sequence $S-S_{\mathrm{L}}-S$ along the solid red line in Fig.~\ref{fig6}c.

To support our interpretation, we show in Figs.~\ref{fig6}d-h $I_\mathrm{c}$, $I_0$ and subgap $I-V_{\mathrm{sd}}$ and $dI/dV_{\mathrm{sd}}-V_{\mathrm{sd}}$ data taken with $V_{\mathrm{gL}}$ swept along these solid lines of corresponding color at different instances of the GS transition; $V_{\mathrm{gR}}$ is also swept to compensate for cross-capacitance. $I_\mathrm{c}$ and $I_\mathrm{0}$ traces (in inset) in Fig.~\ref{fig6}d behave similarly to NRG calculations of $\abs{I_\mathrm{c}}$ in Fig.~\ref{fig5}d. Asymmetric peaks in the blue trace come together in the intermediate green trace and merge into a single broad resonance in the red trace as the coupling $\Gamma_\mathrm{L}$ is increased by tuning the $V_{\mathrm{g1}}$. The asymmetric peaks in $I_\mathrm{c}$ are consistent with a phase-shift of $0-\pi-0$ added to the current-phase relationship when the GS parity changes as even-odd-even. In turn, the culminating single-broad peak is compatible with an enhanced supercurrent from screening of the spin of the left QD, as seen in earlier S-QD-S devices~\cite{kim2013transport,cleuziou2006carbon,Maurand2012}. The absence of two asymmetric peaks in this trace is related to the even parity of the GS for all the relevant charge states.

Figures \ref{fig6}e,f compare the evolution of low-bias $I-V_{\mathrm{sd}}$ traces in gate voltage taken along solid lines of corresponding color in Figs.~\ref{fig6}a,c, to Figs.~\ref{fig6}g,h, which show $dI/dV_{\mathrm{sd}}$ colormaps vs.~$V_{\mathrm{gL}}$ taken along the same respective gate trajectories, but with $V_{\mathrm{sd}}$ swept along a larger bias window within $\pm 2\Delta$, so as to capture the behaviour of YSR subgap $dI/dV_{\mathrm{sd}}$ peaks. 
In Fig.~\ref{fig6}g, the lowest-lying, gapped peaks exhibit a small split-loop structure with kinks at -3.05 V and -3.1 V, which align in gate voltage to peaks in $I_\mathrm{0}$ in the $I-V_{\mathrm{sd}}$ traces from Fig.~\ref{fig6}e, as predicted by our NRG calculations above, and as previously observed in S-QD-S Josephson junctions~\cite{kim2013transport}. The kinks vanish in Fig.~\ref{fig6}h, indicating absence of parity changes. Instead, the peaks exhibit a smooth point of inflexion at $V_{\mathrm{gL}}$=-3.43 V which aligns with a broad resonance in the $V_{\mathrm{gL}}$ dependence of $I_\mathrm{0}$ in the $I-V_{\mathrm{sd}}$ traces from Fig.~\ref{fig6}f.

As in our simple model in Figs.~\ref{fig5}e,f, the lowest-lying peaks appear split in bias voltage by a gapped region~\cite{lee2014spin,jellinggaard2016tuning,deacon2010tunneling,pillet2013tunneling,lee2017scaling,Li2017Jan}, are followed by NDC \cite{kim2013transport,pillet2010andreev,grove2009superconductivity,eichler2007even,lee2012zero,He2020Aug}, and ascribe to the expected shape in gate voltage~\cite{lee2014spin,deacon2010tunneling,Gramich2015Nov,EstradaSaldana2020Jul}. However, closer inspection reveals additional complexity in
the form of reduced gap values and peak replicas, which we will discuss in some detail. If the Bardeen model in Section III is used to interpret the bias position of the lowest-lying gapped peak as $\Delta^*_\mathrm{R}+E_\mathrm{L}$, the measurements in Figs.~\ref{fig9}c,d in Appendix A show that the superconducting gap is effectively reduced with respect to that of the parent Al superconductor ($\Delta$) and that it is different for the left ($\Delta^*_\mathrm{L}=0.110$ meV) and right ($\Delta^*_\mathrm{R}=0.140$ meV) leads. By an effectively reduced gap, we mean that the gapped region in the weak tunnelling regime at $V_{\mathrm{sd}}<2\Delta$ observed at $V_{\mathrm{bg}}=-15$ V in Fig.~\ref{fig9}c is replaced in Fig.~\ref{fig9}d, at the backgate voltage of operation of the device in shell X, $V_{\mathrm{bg}} \approx 10$ V, by conductance resonances above $V_{\mathrm{sd}}>\Delta^*_\mathrm{L}+\Delta^*_\mathrm{R}$.
In electrostatic simulations of gating in InAs nanowire/Al hybrids, reduction of the effective induced gap in the nanowire occurred at more positive gate voltages due to weaker semiconductor/superconductor hybridization, which could explain our measurement~\cite{Antipov2018Aug}.

Additionally, the colormaps display significantly less conducting, bias-symmetric replicas at $V_{\mathrm{sd}}>\Delta^*_\mathrm{R}+E_\mathrm{L}$, as well as one or more faint replicas at $V_{\mathrm{sd}}<\Delta^*_\mathrm{R}+E_\mathrm{L}$, the lowest of which can cross zero bias at parity crossings (see Appendix C). These replicas are often followed by NDC. In addition to replicas, on occasion a few straight horizontal lines cross the colormaps, such as in Fig.~\ref{fig6}h, which shows a pair of bias-symmetric lines at $V_{\mathrm{sd}}=\Delta^*_\mathrm{R}$. The colormaps also display additional lines with opposite curvature to that of the lowest-lying gapped peaks; a clear example of this is seen in Fig.~\ref{fig6}h around $V_{\mathrm{gL}}=-3.43$ V as split-loops terminating at the main YSR peaks. These additional lines can also exhibit replicas. The intricate replica behavior has been previously related to multiple peaks inside the superconducting gap of the hybrid nanowire-superconductor leads~\cite{su2018mirage,deng2016majorana}, to multiple Andreev reflection~\cite{nilsson2011supercurrent,Villas2020Jun}, quasiparticle relaxation, thermally excited transport~\cite{Ruby2018Apr,Huang2020Jul,Kumar2014Feb} and/or to inelastic Andreev tunnelling~\cite{Gramich2015Nov}. Understanding the origin of these replicas, some of which may be visible in this work due to the unprecedented resolution of our data (the full width at half maximum of YSR peaks can be as low as 10 $\mu$V, as shown in Fig.~\ref{fig14}b), is outside of the scope of this work. In what follows, we will focus only on the gate dependence of the curvature of the lowest-lying pair of gapped peaks, and on its relation to $I_\mathrm{0}$.

We now bring shell X back to the honeycomb regime, this time with gate settings which put it close to a GS transition due to YSR screening of the right-QD spin. Fig.~\ref{fig7}a shows a colormap of zero-bias conductance which represents the charge stability diagram of shell X in this regime. The sequence of colormaps in Figs.~\ref{fig7}a-f shows tuning of the stability diagram from the honeycomb pattern into a pattern lacking parity lines from the right QD. We interpret the change as stemming from an effective increase in $\Gamma_\mathrm{R}$ due to our tuning of the rightmost gate voltage, $V_{\mathrm{g5}}$. The change is more subtle as $V_{\mathrm{gL}}$ also affects $\Gamma_\mathrm{R}$, as mentioned above. Due to this, the upper section of the stability diagram transitions into a new GS faster than the lower one. We are able to model this qualitatively by introducing a linear dependence of $\Gamma_\mathrm{R}$ on $n_\mathrm{L}$ within a zero-bandwidth approximation of the superconducting continuum~\cite{grove2017yu}, as shown in Appendix F, and find that this is a faithful demonstration of a $S_{\mathrm{00}}$ to $D_{\mathrm{01}}$ GS transition in the (1,1) charge sector, and of a $D_{\mathrm{R}}$ to $S_{\mathrm{1}}$ GS transition in the (0,1) and (2,1) charge sectors. In Appendix F, we also show that GS transitions obtained by increasing $\Gamma_\mathrm{L}$, as those shown Fig.~\ref{fig6}, are not significantly affected by introducing a dependence of $\Gamma_\mathrm{R}$ on $n_\mathrm{L}$.

Finally, departing from the pattern in Fig.~\ref{fig7}f, in which the right-QD spin of DQD-shell X is screened, we demonstrate simultaneous screening of the left-QD spin, and therefore, full YSR screening of the spins of shell X. Figure \ref{fig8}a-c shows the step-by-step modification of this stability diagram, which contains only parity lines of the right QD in shell X in Fig.~\ref{fig8}a, to a diagram lacking parity lines of this shell in Fig.~\ref{fig8}c, which marks a $D_{\mathrm{01}}$ to $S_{\mathrm{11}}$ GS transition. In Fig.~\ref{fig8}c, after the GS transition occurs, the charge stability diagram of shell X shows a broad peak of conductance as the remains of the larger availability of conduction channels for Cooper pair transport in the (1,1) charge sector. The GS transition is also reflected in the gate dependence of $I_\mathrm{c}$, $I_\mathrm{0}$ and YSR peaks of largest conductance in Figs.~\ref{fig8}d-h, obtained along gate paths given by solid lines of corresponding color in Figs.~\ref{fig8}a-c, which cross the (1,1) charge sector. As in Fig.~\ref{fig6}d, Fig.~\ref{fig8}d shows that $I_\mathrm{c}$ and $I_\mathrm{0}$ (in inset) gradually change from a split-peak structure in the blue trace to a broad peak in the red trace. As in Figs.~\ref{fig6}e-h, parity changes manifested in Fig.~\ref{fig8}g in two kinks in the split-loop YSR structure which align to $I_\mathrm{0}$ peaks in $I-V_{\mathrm{sd}}$ traces in Fig.~\ref{fig8}e are replaced by an all-singlet parity sequence, which expresses itself in the absence of kinks in Fig.~\ref{fig8}h, and in the smooth inflection point at $V_{\mathrm{gL}}=-1.9$ V which coincides in Fig.~\ref{fig8}f with a broad $I_\mathrm{0}$ peak in $I-V_{\mathrm{sd}}$ traces.

From the succession of Figs.~\ref{fig7}a-f and Figs.~\ref{fig8}a-c, we can see that the GS of the (1,1) charge sector, which started in Fig.~\ref{fig7}a as an inter-dot singlet, $S_{\mathrm{00}}$, has been gate-tuned in Fig.~\ref{fig8}c into independently screened YSR singlets, $S_{\mathrm{11}}$. This GS transition occurred through an intermediate YSR doublet phase, $D_\mathrm{{01}}$, as shown in Figs.~\ref{fig7}f and \ref{fig8}a.

\section{Conclusion}

In summary, we have demonstrated one- and two-impurity YSR physics in a shell of a DQD hybrid nanowire. We obtain the GS from the gate dependence of $I_\mathrm{c}$, of $I_\mathrm{0}$ and of the gate dispersion of YSR subgap conductance peaks, supported by the step-by-step tuning of the stability diagram to different endpoints of the phase diagrams. We find a reasonable qualitative agreement between experiment and theory. However, technical difficulties which prevent the measurement of $\Gamma_L$ and $\Gamma_R$, together with cross-couplings in the device, have as a result that the S-DQD-S model cannot be quantitatively established in relation to the experiment without free fitting parameters. Quantitative comparison between experiment and theory is also complicated by lack of independent measurement of $R$ within our model for the characteristic of the JJ circuit, which leads to extraction of $I_\mathrm{c}$ only up to a factor $\sqrt{\gamma} I_\mathrm{c}$, and by the complexity of the reduced hybrid nanowire-superconductor gap, which is reflected in multiple replicas of the YSR state filling the gap between the YSR state and the parent Al superconducting gap edge.

As a spectroscopic probe of parity changes, the narrow zero-bias conductance peak of the effectively voltage-biased Josephson junction~\cite{Naaman2001Aug,Rodrigo2004Aug,Randeria2016Apr,Cho2019Jul,Peters2020Jul,Liu2020} maintains the sharpness of the relevant features in all charge, or more accurately, parity stability diagrams independently of the tunnelling rates. This is in stark contrast to the case of magnetic impurities (spinful QDs) coupled to normal metals, which broaden the conductance features at strong hybridization~\cite{jeong2001kondo,bork2011tunable,chorley2012tunable,spinelli2015exploring,chang2009kondo,georges1999electronic}. The zero-bias conductance peak also provides very direct access to $I_\mathrm{c}$, just as it provides access to the local superfluid density in scanned Josephson-tunneling microscopy, which has been used recently to detect Cooper-pair density waves on surfaces of Bi$_{2}$Sr$_{2}$CaCu$_{2}$O$_{8+x}$ and NbSe$_2$~\cite{Hamidian2016,Liu2020}.
Unlike scanning tunnelling spectroscopies of dimers of magnetic adatoms on superconducting surfaces \cite{Ji2008Jun,Ruby2018Apr,Kamlapure2018Aug}, our DQD realization comprises two spin-1/2 states which are completely screened by individual superconducting channels. The Kondo-YSR analogy breaks down towards zero temperature, as confirmed by the existence of doublet domains in the phase diagram at $k_BT \ll \Delta$.

By virtue of $t_\mathrm{d}$, non-magnetic, gap-protected superpositions of the two singlet states found, the exchange singlet $S_{\mathrm{00}}$ and the independently screened YSR singlet $S_{\mathrm{11}}$, could be prepared in future works for parameters close to the anti-crossing in the phase diagram of Fig.~\ref{fig3}~\cite{Wineland2013Jul}, with the purpose of using them as qubits~\cite{DiVincenzo2000Sep}.
In addition, the demonstrated gate control of a two-site quantum dot chain in superconducting proximity is a crucial step towards the implementation in our hybrid wires of the YSR analog of Doniach's Kondo necklace \cite{Doniach1977} and of the Kitaev chain \cite{kitaev2001unpaired,sau2012realizing,fulga2013adaptive}, complementing ongoing research of emergent manifestations of topology \cite{Mourik2012,Nadj-Perge2014Oct,Ruby2015Nov,Pawlak2016Nov,deng2016majorana,Grivnin2019Apr,Manna2020Apr}.

\medskip 
\textbf{Data availability.}
All data needed to evaluate the conclusions in the paper are present in the paper. Raw data used to produce the experimental figures in the paper can be found at the repository ERDA of the University of Copenhagen at \url{https://doi.org/10.17894/ucph.c83b3677-34ff-4d10-8f5a-36785d1ed59e}.

\begin{acknowledgments}
We thank A. Jellinggaard, M.C. Hels, and J. Ovesen for experimental assistance. The project received funding from the European Union’s Horizon 2020 research and innovation program under the Marie Sklodowska-Curie grant agreement No.~832645. We additionally acknowledge financial support from the Carlsberg Foundation, the Independent Research Fund Denmark, QuantERA ’SuperTop’ (NN 127900), the Danish National Research Foundation, Villum Foundation project No.~25310, and the  Sino-Danish Center. P.~K. acknowledges support from Microsoft and the ERC starting Grant No.~716655 under the Horizon 2020 program. R.~\v{Z}. acknowledges support from the Slovenian Research Agency (ARRS) under Grants No.~P1-0044 and J1-7259.
\end{acknowledgments}

\section*{Appendix A: Extraction of DQD parameters}

\begin{figure*} [t!]
\includegraphics[width=0.85\linewidth]{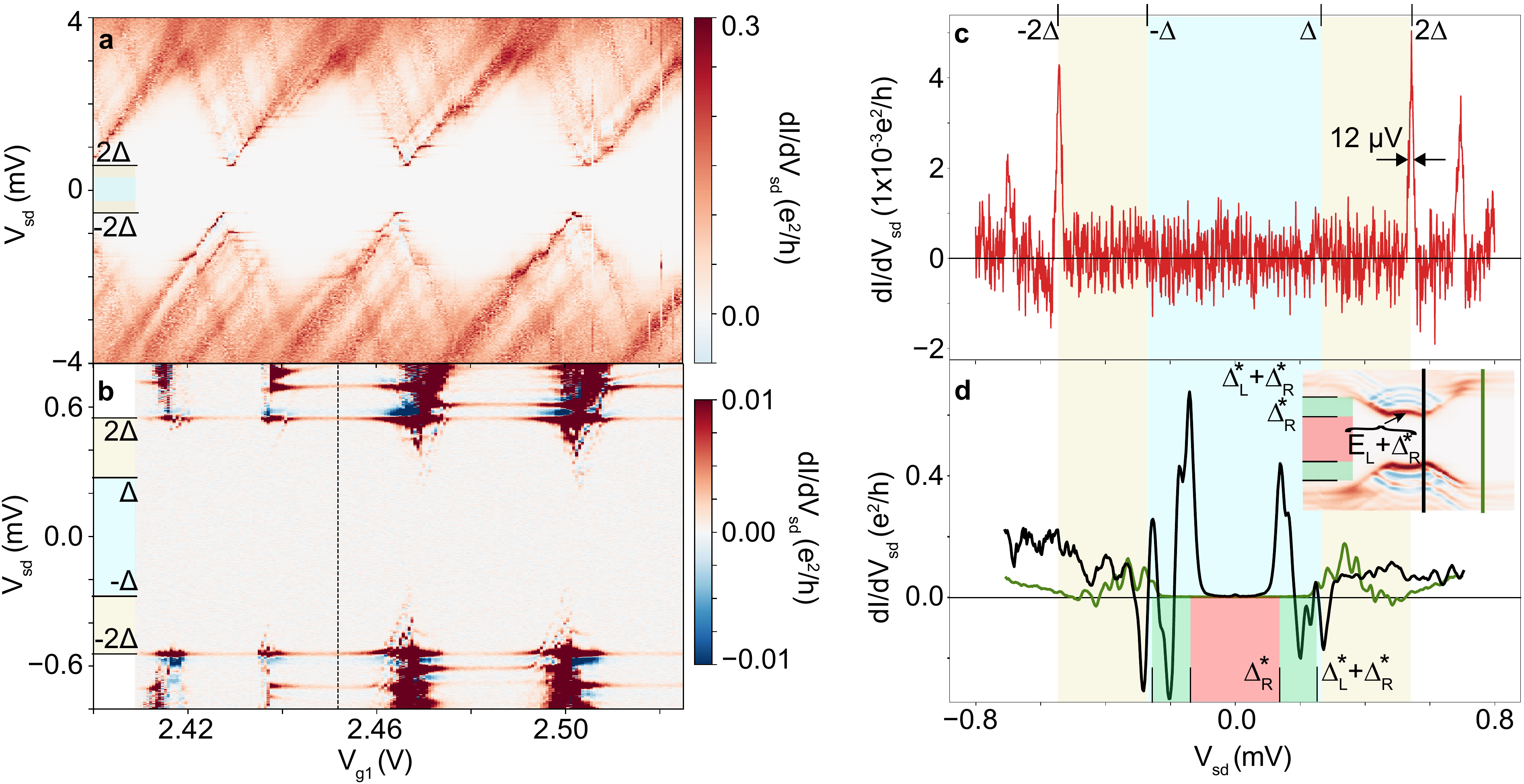}
\caption{(a) $dI/dV_{\mathrm{sd}} (V_{\mathrm{sd}},V_{\mathrm{g1}})$ colormap taken with other gates set at $V_{\mathrm{g2}}=2.25$ V, $V_{\mathrm{g3}}=-6.5$ V, $V_{\mathrm{g4}}=-2.16$ V, $V_{g5}=-6.5$ V and $V_{\mathrm{bg}}=-15$ V. (b) $dI/dV_{\mathrm{sd}} (V_{\mathrm{sd}},V_{\mathrm{g1}})$ colormap measured along the same gate trajectory as (a) and zoomed into the superconducting gap. A small gate shift occurred at $V_{\mathrm{g1}} \approx 2.438$ V. In (a,b), the colorscale has been saturated to highlight faint features in the data. (c) $dI/dV_{\mathrm{sd}} (V_{\mathrm{sd}})$ trace obtained along the dashed line in (b). (d) $dI/dV_{\mathrm{sd}} (V_{\mathrm{sd}})$ traces measured along line of same color in the inset colormap, which corresponds to the same colormap as the one shown in Fig.~\ref{fig6}g. Only data in (d) corresponds to shell X. Blue and yellow shadings separate $\Delta, 2\Delta$ regions, while red and green shadings separate $\Delta^*_\mathrm{R}, \Delta^*_\mathrm{L}+\Delta^*_\mathrm{R}$ regions.
}
\label{fig9}
\end{figure*}

\begin{figure*} [t!]
\includegraphics[width=0.9\linewidth]{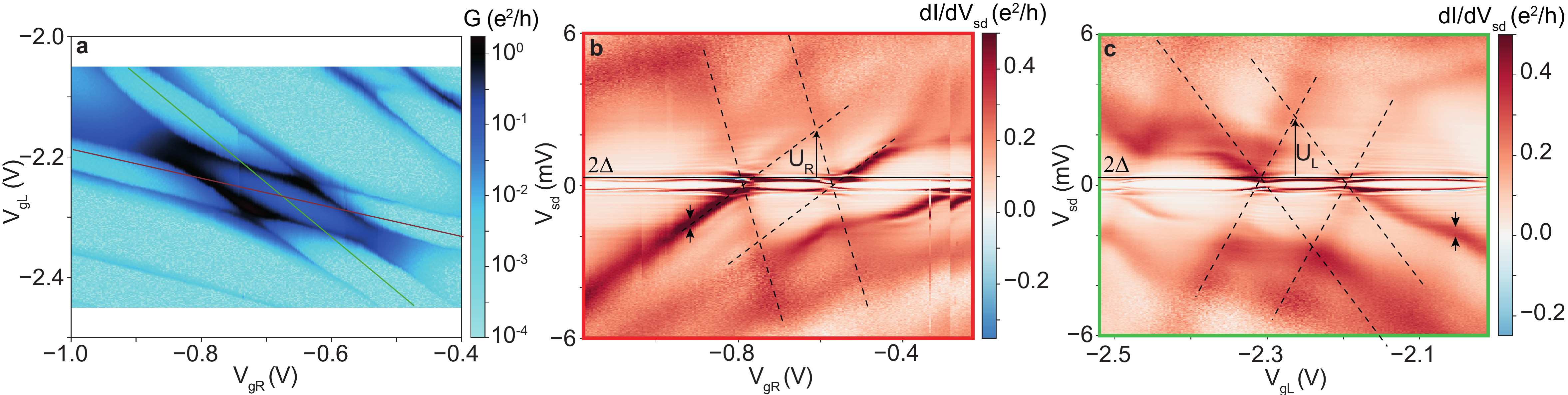}
\caption{(a) Colormap of zero-bias conductance, $G$, vs.~$(V_{\mathrm{gL}},V_{\mathrm{gR}})$ for shell X measured with other gates set at $V_{\mathrm{g1}}=-9.2$ V, $V_{\mathrm{g3}}=-9.3$ V, $V_{\mathrm{g5}}=-0.25$ V and $V_{\mathrm{bg}}=11$ V. (b,c) Colormaps of $dI/dV_{\mathrm{sd}}-V_\mathrm{{sd}}$ vs.~gate voltage representing Coulomb-diamond spectroscopy of the (b) right QD, (c) left QD. Gate trajectories followed in each colormap are indicated by red and green lines in (a), respectively; however, for simplicity, only (b) $V_{\mathrm{gR}}$ and (c) $V_{\mathrm{gL}}$ gates are indicated on the horizontal axes. Dashed lines are guides to the eye to follow Coulomb diamonds. In (b) and (c), the AC lock-in excitation is set to 20 $\mu$V.
}
\label{fig10}
\end{figure*}

\begin{figure} [t!]
\includegraphics[width=1\linewidth]{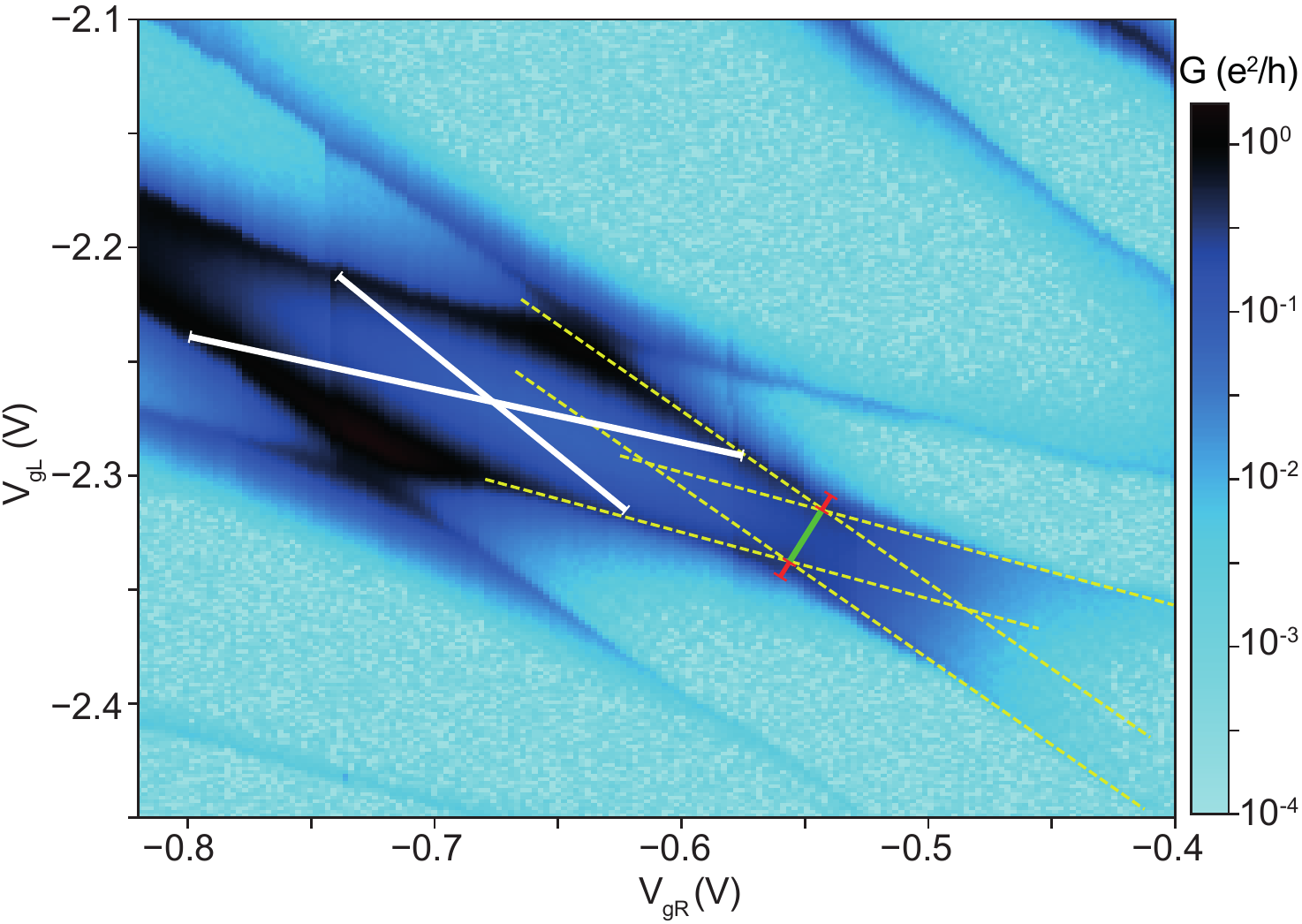}
\caption{Zoomed-in version of the colormap of zero-bias conductance vs.~$(V_{\mathrm{gL}},V_{\mathrm{gR}})$ for shell X shown in Fig.~\ref{fig10}. Lines are used to estimate $t_\mathrm{d}$ and $U_\mathrm{d}$ as described in Appendix A.  
}
\label{fig11}
\end{figure}

To measure $\Delta$, we deplete the device by setting $V_{\mathrm{bg}}=-15$ V (which is $\approx 25$ V more negative than $V_\mathrm{{bg}}$ used in the measurements on shell X in the main text) and perform Coulomb-diamond spectroscopy as shown in Fig.~\ref{fig9}a. The repetitive Coulomb-blockade pattern is indicative of a multilevel QD regime. The Coulomb diamonds appear split by a forbidden bias window. Zooming in on this window, as shown in Fig.~\ref{fig9}b, a pair of bias-symmetric lines which cross all diamonds is identified at $eV_{\mathrm{sd}}=\pm 2\Delta$. This feature stems from the BCS coherence peak from the left lead at energy $-\Delta$ probing the coherence peak in the right lead at energy $+\Delta$ via co-tunnelling through the QDs. At parity crossings, which correspond to the apexes of split Coulomb diamonds in this opaque regime, a few pairs of bias-symmetric subgap states come down, the lowest of which reaches exactly $\Delta$. This occurs when the coherence peak at $\Delta$ in the right lead probes a YSR excitation in the left lead which crosses zero energy at parity crossings. At $e|V_{\mathrm{sd}}|<\Delta$, the conductance is blocked at all $V_{\mathrm{g1}}$ values. In Fig.~\ref{fig9}c, we plot a linecut at the center of one Coulomb diamond to display the position of the quasiparticle tunnelling peaks at $2\Delta$ and their extremely narrow width~\cite{Huang2020Jul}. We measure $2\Delta=0.54$ meV, which provides a value of $\Delta$ identical to the one obtained from S-QD-N devices made from the same batch of nanowires~\cite{EstradaSaldana2020Jul}. The position of horizontal lines outside $2\Delta$ depends on the Coulomb diamond chosen, as can be seen in Fig.~\ref{fig9}b; we ascribe these lines to the probing of QD states in co-tunnelling by the sharp superconducting singularities. The trace in Fig.~\ref{fig9}c is compared to two traces in Fig.~\ref{fig9}d displaying YSR excitations in shell X which are extracted from Fig.~\ref{fig6}g to show graphically that $\Delta^*_\mathrm{R}<\Delta$ and that $\Delta^*_\mathrm{L}+\Delta^*_\mathrm{R}<2\Delta$, where $\Delta^*_\mathrm{L}$ and $\Delta^*_\mathrm{R}$ are the effectively reduced gaps in the left and right leads at the backgate voltage at which the device is operated, as in Section IV. If we assign a bias position $E_\mathrm{L}+\Delta^*_\mathrm{R}$ to the innermost gapped YSR peaks, we measure $\Delta^*_\mathrm{R}=0.140$ meV from the black trace, which is taken at a parity crossing where $E_\mathrm{L}=0$. In turn, we find $\Delta^*_\mathrm{L}+\Delta^*_\mathrm{R}=0.250$ meV from the green trace, which is taken in the (0,2) charge state, where $E_\mathrm{L}$ approaches asymptotically to $\Delta^*_\mathrm{L}$. From this, we deduce $\Delta^*_\mathrm{L}=0.110$ meV.

To measure $U_\mathrm{L}$, $U_\mathrm{R}$, we perform Coulomb-diamond spectroscopy on each QD in shell X as shown in Fig.~\ref{fig10}. In spite of the larger couplings of the DQD in this regime, we observe faded Coulomb diamonds from which we can trace the actual diamonds as denoted by dashed lines. $U_\mathrm{L}$, $U_\mathrm{R}$ correspond to the bias difference between the apex of the central Coulomb diamond and the edge of the superconducting gap at $2\Delta$. An indication of the order of magnitude of the tunnelling rates of the DQD can be obtained from the full-width-at-half-maximum of Coulomb lines outside the gap between the pairs of arrows in Figs.~\ref{fig10}b,c, which is of 0.6 meV in both cases.

To estimate $t_\mathrm{d}$ and $U_\mathrm{d}$, we use the curvature of the parity lines at the (1,1), (0,2) charge transition in the charge stability diagram of shell X. We first prolong the parity lines of the left and right QDs, as indicated by the yellow dashed lines in Fig.~\ref{fig11}. We join their two intersections by a green line which separates the (1,1) and (0,2) charge sectors. Afterwards, by projecting the green line into the two white lines which indicate the charging energies of the QDs, we convert the gate scale of the green line into energy. The red lines along the green line indicate the distance from the curved parity lines in the colormap to each of the two intersections of the yellow dashed lines and are each equal to $2t_\mathrm{d}$; the green line summed to the two red lines is $\sqrt{2}U_\mathrm{d}+4 t_\mathrm{d}$. This procedure assumes that the parity stability diagram in the superconducting state is similar to that of the normal state.
In this way, we measure $t_\mathrm{d} \approx 0.03-0.05$ meV and $U_\mathrm{d} \approx 0.13-0.23$ meV.
 
\section*{Appendix B: Model of the circuit}

In this appendix we elaborate on the extended RCSJ model necessary to explain the $I-V_{\mathrm{sd}}$ and $dI/dV_{\mathrm{sd}}-V_{\mathrm{sd}}$ curves observed in the experiment. The extension, compared to standard RCSJ, consists of an additional series resistance $R$ and shunt capacitance $C$. This model has been applied to ultra-small Josephson junctions~\cite{Vion1996Oct} and S-QD-S setups~\cite{Jorgensen2007}. Disregarding $C_{\mathrm{filter}}$ in the circuit depicted in Fig.~\ref{fig2}c, Kirchhoff's laws and the Josephson relation yield two coupled Langevin equations:
\begin{align}
\frac{du}{d\tau}&=\frac{1}{\alpha}\left(\frac{V}{R_s I_c}-\sin \phi-\frac{R}{R_s}(u+L(\tau))-\frac{1}{\alpha_0}\frac{d^2\phi}{d\tau^2}\right) \label{ueq} \\[0.2cm]
\frac{d\phi}{d\tau}&=u+L(\tau)-\sin \phi-\frac{1}{\alpha_0}\frac{d^2\phi}{d\tau^2}.
\end{align}
Here $u=V_{\mathrm{sd}}/RI_\mathrm{c}$ denotes the dimensionless voltage across the shunt capacitor $C$ and we have introduced the dimensionless time $\tau=\omega_{RL}t$ with $\omega_{RL}=R/\textup L$, where $\textup L=\Phi/I_\mathrm{c}$ is the self-inductance of the Josephson junction set by the critical current, $I_\mathrm{c}$, and $\Phi=\hbar/2e$. The two different $RC$ frequencies, $\omega_{RC_{\mathrm{J}}}=(RC_{\mathrm{J}})^{-1}$ and $\omega_{RC}=(RC)^{-1}$, enter these equations via the dimensionless parameters $\alpha=\omega_{RL}/\omega_{RC}=R^2CI_c/\Phi$ and $\alpha_0=\omega_{RC_{\mathrm{J}}}/\omega_{RL}=\Phi/R^2 C_\mathrm{J} I_\mathrm{c}$. Finally, $L(\tau)$ is a dimensionless stochastic parameter describing white-noise voltage fluctuation of $V_{\mathrm{sd}}$. $L(\tau)$ is composed of Nyquist noise from the series resistor ($R$) at temperature $T^*$ and a stray voltage noise characterized by a variance, $K$, to account for imperfect filtering, such that $\langle L(\tau)L(\tau+\Delta \tau)\rangle = \left(\frac{2k_B T^*}{\Phi I_c}+\frac{K}{RI_c\Phi}\right)\delta(\Delta\tau)$. Noise from the source resistance $R_\mathrm{s}$ is assumed to be excluded by the RC filters. As $K$ and $T^*$ can not be independently measured in this setup we ascribe all noise to the series resistor ($R$) with an effective temperature $T=T^*+K/(2Rk_B)$ which can be different from the fridge temperature.

For $\alpha\gg 1$, $u$ changes slowly with time, which allows us to solve for $\phi$'s equilibrium distribution keeping $u$ constant. If furthermore $\alpha_{0}\gg 1$, solving the stochastic equation for the $\phi$-distribution is equivalent to the original RCSJ problem~\cite{Anchenko1969} with the solution,
\begin{equation}\label{ivan}
\langle \sin\phi\rangle = \operatorname{Im}\frac{I_{1-iuA}(A)}{I_{-iuA}(A)}
\end{equation}
where $A=\Phi I_\mathrm{c}/k_{B}T$ and $I_n(z)$ is a Bessel function. Inserting this solution into Eq.~\eqref{ueq} a constant average voltage, $\dot{\langle u\rangle}=0$, may be enforced by satisfying the equation
\begin{align}
\frac{V}{R_\mathrm{s} I_\mathrm{c}}-\frac{R}{R_\mathrm{s}}\langle u \rangle=\operatorname{Im}\frac{I_{1-i\langle u\rangle A}(A)}{I_{-i\langle u\rangle A}(A)}\label{VolCurr}.
\end{align}
Two limits of $R/R_{\mathrm{s}}$ are of interest. For small $R/R_{\mathrm{s}}$, depending on $I_\mathrm{c}$ and $T$, this equation has three solutions for $\langle u\rangle$ arising from the non-monotonic behavior of the right-hand side, and the setup is effectively current biased, exhibiting hysteretic $\langle I\rangle-V$ curves. For large $R/R_{\mathrm{s}}$, the solution for $\langle u\rangle$ is unique and the system is effectively voltage biased and can exhibit NDC. The latter case is consistent with the experiment.

\begin{figure} [b]
\includegraphics[width=1\linewidth]{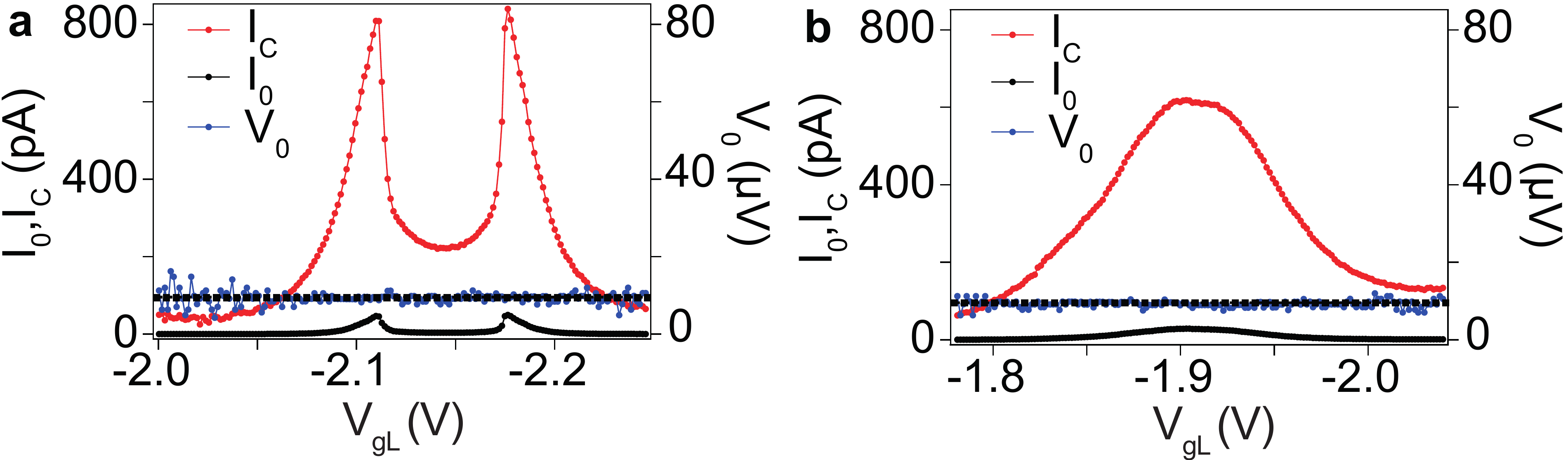}
\caption{(a,b) Fitted $I_\mathrm{c}$, extracted cusp current, $I_\mathrm{0}$ (left axes), and extracted cusp voltage, $V_\mathrm{0}$ (right axes) from $I-V_{\mathrm{sd}}$ data from Figs.~\ref{fig8}e,f, respectively. Data from all measured $I-V_{\mathrm{sd}}$ traces in the gate range indicated is included, and not just every third trace. An horizontal dashed line corresponds to $V_\mathrm{0}=9.1$ $\mu$V. Deviations from constant $V_\mathrm{0}$ occur only for small current ($I_\mathrm{0}<0.3$ pA); e.g., between $V_{\mathrm{gL}}=-2$ and -2.05 V, where data noise prevents an accurate extraction of $V_\mathrm{0}$.
}
\label{fig12}
\end{figure}

Experimental $I-\langle V_{\mathrm{sd}}\rangle$ traces can now be fitted to
\begin{equation}
\langle I \rangle=
I_c\operatorname{Im}\frac{I_{1-i\langle u \rangle A}(A)}{I_{-i\langle u \rangle A}(A)},
\label{Ivanchen}
\end{equation}
which is identical to an earlier formula derived by Ivanchenko and Zil'berman~\cite{Anchenko1969}, while the derivative of this formula with respect to $\langle V_{\mathrm{sd}}\rangle$ allows fitting of $dI/dV_{\mathrm{sd}}-V_{\mathrm{sd}}$ traces. In a typical Josephson junction $A\gg1$ and the cusp voltage  scale is set by $V_\mathrm{0}\approx R I_\mathrm{c}$. In our measurements $V_\mathrm{0}$ is constant over a wide range of cusp current, $I_\mathrm{0}$, as we show in Fig.~\ref{fig12} where extracted $I_\mathrm{0}$, $V_\mathrm{0}$ and fitted $I_\mathrm{c}$ is shown for two linecuts in the experiment.

This is consistent with the large-noise limit, $A<1$, of Eq.~(\ref{Ivanchen}) where one can perform a Taylor expansion of Eq.~(\ref{Ivanchen}) around $A\approx 0$ to obtain~\cite{Anchenko1969},
\begin{equation}
\langle I\rangle \approx \frac{1}{2}\frac{I_c A^2 \langle u\rangle }{1+A^2\langle u\rangle^2}
\label{SimpleIvan}
\end{equation}
from which it is clear that the cusp voltage $V_\mathrm{0}=\operatorname{arg}\max \langle I\rangle(\langle V_{sd}\rangle) = Rk_BT/\Phi$ is solely determined by noise and is independent of $I_\mathrm{c}$. Cusp current, 
\begin{equation}
I_\mathrm{0}=\langle I\rangle(V_\mathrm{0})=I_cA/4
\end{equation}
and zero-bias conductance, 
\begin{equation} 
G=d\langle I \rangle/d\langle V_{\mathrm{sd}}\rangle(0)=A^2/2R
\end{equation}
can also easily be identified in this limit. We also identify a scaling $\gamma$ with $\Bar{R}=R/\gamma$, $\Bar{T}=T\gamma$ and $\Bar{I}_\mathrm{c}=\sqrt{\gamma}I_\mathrm{c}$ which keeps $V_\mathrm{0}$ and Eq.~(\ref{SimpleIvan}) unchanged.

\begin{figure} [t]
\includegraphics[width=1\linewidth]{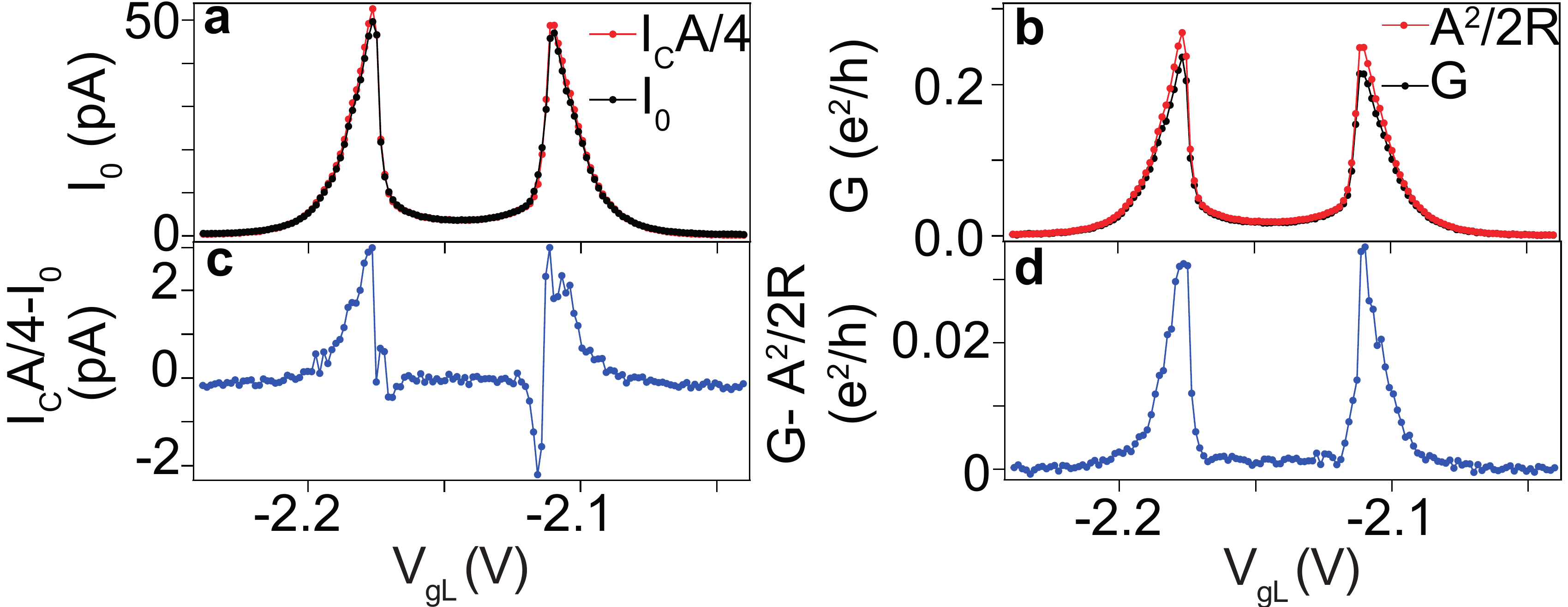}
\caption{(a-d) Comparison between the gate dependence of rescaled $I_\mathrm{c}$ and (a) measured $I_\mathrm{0}$ from Fig.~\ref{fig14}c, (b) $G$ linecut through Fig.~\ref{fig14}d. To obtain $I_\mathrm{0}$ in (a), data from all measured $I-V_{\mathrm{sd}}$ traces corresponding to the gate trajectory in \ref{fig14}c is included, and not just every third trace. (c,d) Difference between the scaled and measured curves in (a,b).}
\label{fig13}
\end{figure}

If the limits above are satisfied we indeed have a model where: 1) The junction is overdamped and non-hysteretic, 2) the junction is effectively voltage biased allowing measurements of NDC and 3) the $I-V_{\mathrm{sd}}$ curves are cusped with an $I_\mathrm{0}$-independent cusp voltage $V_\mathrm{0}$. We will now discuss the validity of the above limits. By fixing $T=80$ mK, consistent fits of Eq.~(\ref{Ivanchen}) can be done for all gate ranges yielding a constant series resistance $R=3$ k$\Omega$ and variable $I_\mathrm{c}$ in the range $I_\mathrm{c}\sim 0.02-3$ nA which roughly corresponds to $\alpha\sim 10-100$ and $\alpha_0\sim 10^4-10^5$ using the geometric estimates of $C$ and $C_\mathrm{J}$, consistent with our assumption that $\alpha,\alpha_{0}\gg 1$. For these parameters Eq.~(\ref{VolCurr}) has only one solution, consistent with the junction being voltage biased. Lastly, for these parameters $A\leq 1$ for the fitted range of $I_\mathrm{c}$ with $A\approx 1$ for the largest $I_\mathrm{c}$. In Fig.~\ref{fig13}, we compare the simple large-noise-limit expressions for $I_\mathrm{0}$ and $G$ using $I_\mathrm{c}$ from the fitting with measured $I_\mathrm{0}$ and $G$ and find excellent agreement with the largest discrepancies occurring for large $I_\mathrm{c}$ values consistent with our Taylor expansion in $A\propto I_\mathrm{c}$.

As the above analysis is a fully classical treatment we will now discuss the impact of quantum fluctuations.
The characteristic frequencies of the circuit are estimated to be $\omega_{RC}\sim 0.01$ GHz, $\omega_{RL}\sim 0.1-10$ GHz, and $\omega_{RC_{\mathrm{J}}}\sim 10^{5}$ GHz. With $T=80$ mK, i.e. $k_{B}T/\hbar\sim 10$ GHz, and $I_\mathrm{c}=0.02-3$ nA, we arrive at the following hierarchy of frequencies:
\begin{align}
\omega_{RC}\ll \omega_{RL}\lesssim k_{B}T/\hbar\ll\omega_{RC_{\mathrm{J}}},
\end{align}
For these values, one may safely neglect the large capacitance, $C$, in the effective circuit impedance experienced by the junction
\begin{equation}
Z_{eff}(\omega)=\left(1/R+i C_\mathrm{J} \omega+1/(i\textup L\omega)\right)^{-1},
\end{equation}
and which determines the mean-squared phase fluctuations, $S_{\phi}=\langle(\phi-\langle\phi\rangle)^{2}\rangle$, which is given by Refs.~\onlinecite{Grabert1998Nov,Grabert1999May,Joyez2013May}, 
\begin{equation}
S_{\phi}=\int_{-\infty}^{\infty}\frac{d\omega}{\omega}\frac{\operatorname{ReZ_{\mathrm{eff}}(\omega)}}{2R_Q}\frac{1}{1-e^{-\hbar\omega/k_{B}T}}
\end{equation}
with $R_\mathrm{Q}=h/4e^{2}\approx 6.5$ k$\Omega$. Within a fully quantum mechanical treatment, this correlation function in turn leads to an approximately exponential reduction of the critical current~\cite{Grabert1998Nov,Joyez2013May}:
\begin{align}
I_{\mathrm{c}}^{\ast}\approx I_{\mathrm{c}}\exp(-S_{\phi}). 
\end{align}
As a function of temperature, $S_{\phi}$ is roughly constant and due to quantum fluctuations for $k_{B}T< \hbar\omega_{RL}$. For $\hbar\omega_{RL}<k_{B}T$, corresponding to the parameters found above, the fluctuations are largely classical and $S_{\phi}$ increases linearly with $T$, consistent with the classical treatment employed above. Since $k_{B}T$ is just barely larger than $\hbar\omega_{RL}$, and $R$ just barely smaller than $R_Q$, a further reduction of $I_\mathrm{c}$ due to quantum fluctuations must be expected to reduce the actual values for $I_\mathrm{c}$, which we deduce from the strictly classical analysis above and report in the figures throughout the main text. Nevertheless, the marked gate dependence of $I_\mathrm{c}$ remains and this is what provides the real value of the zero-bias conductance peak as a probe of YSR screening and the concomitant quantum phase transitions reported in this paper.

Summarizing the circuit analysis, the magnitudes of $I_\mathrm{c}$ obtained by fitting $I-V_{\mathrm{sd}}$ and $dI/dV_{\mathrm{sd}}-V_{\mathrm{sd}}$ traces are expected to be slightly overestimated by leaving out quantum, and retaining only thermal fluctuations of the electromagnetic environment. Additionally, $T$ has been fixed by an independent measurement which is only an estimate of the effective temperature experienced by the circuit. Relaxing this constraint on $T$, we can still obtain consistent fits utilizing a global scaling of $R$ and $I_\mathrm{c}$. The value of these measurements therefore lies in the gate dependence of $I_\mathrm{c}$, not its amplitude which can include a global scaling, in good qualitative agreement with NRG calculations.

\section*{Appendix C: Asymmetries in cusp current}

\begin{figure} [t!]
\includegraphics[width=1\linewidth]{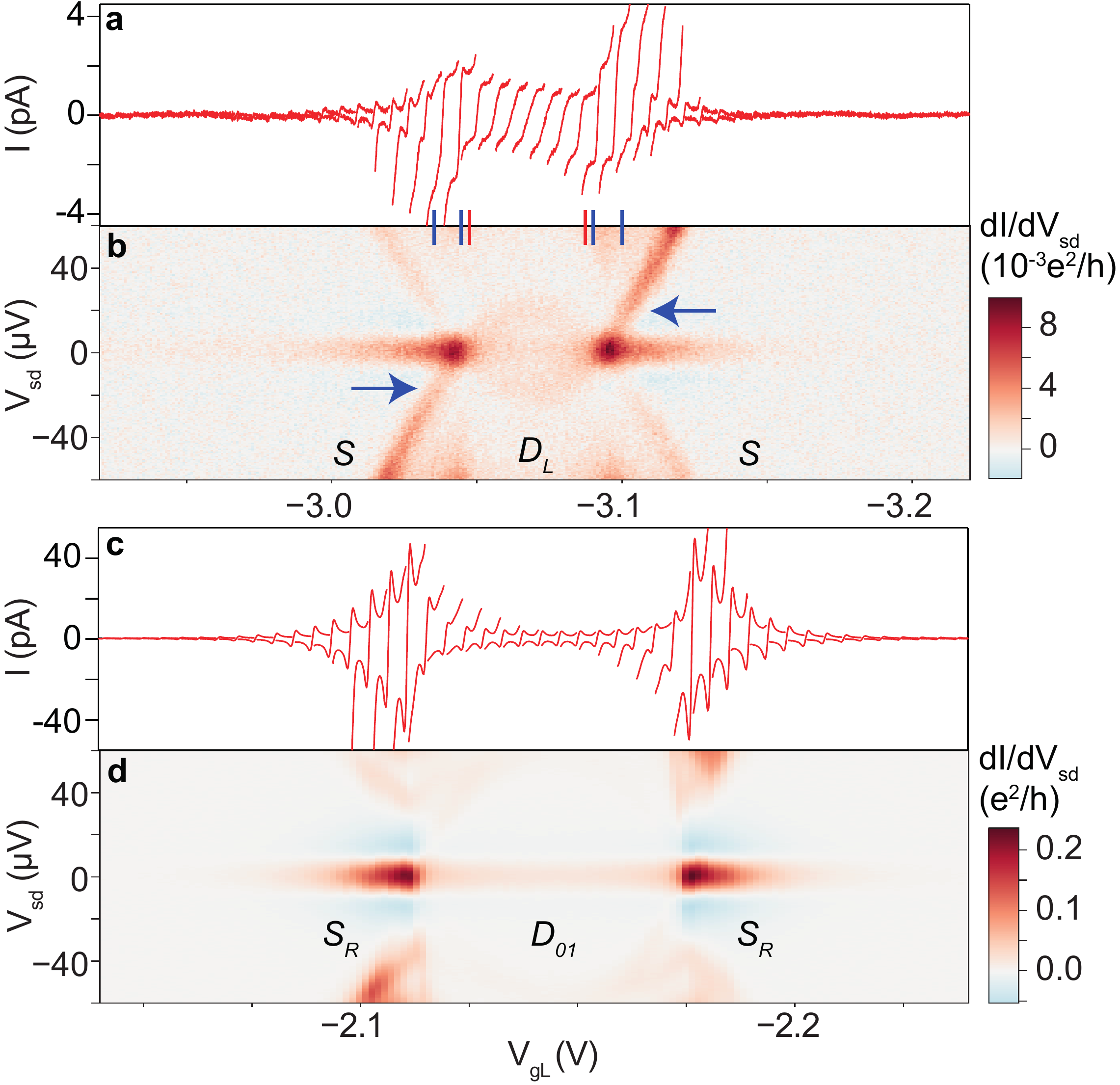}
\caption{(a-d) Comparison between the gate dependence of (a,c) $I-V_{\mathrm{sd}}$ traces and respective (b,d) $dI/dV_{\mathrm{sd}}$ colormaps which represent the gate dependence of the corresponding low-lying non-gapped YSR subgap states for (a,b) the $S-D_{\mathrm{L}}-S$ and (c,d) the $S_{\mathrm{R}}-D_{\mathrm{01}}-S_{\mathrm{R}}$ GS sequences. In (a) (and (c)), to avoid crowding, only every seventh (every third) trace is plotted, and the traces are horizontally shifted by 65 $\mu$V with respect to each other. In each trace, $V_{\mathrm{sd}}$ is swept from $-60$ to $60$ $\mu$V. To acquire the datasets in (a-d), the gates were swept along the solid lines in Fig.~\ref{fig6}a for plots (a,b), and in Fig.~\ref{fig8}a for plots (c,d).}
\label{fig14}
\end{figure}

While the model displays $I_\mathrm{0}$ symmetric in $V_{\mathrm{sd}}$, in some instances the experiment fails to do so, with the discrepancy attributed to conduction through low-lying YSR subgap peaks which approach $V_{\mathrm{sd}}=0$. In Fig.~\ref{fig14} we show two examples in which subgap peaks approach zero bias, and the effect which this has on the symmetry of $I_\mathrm{0}$ in $V_{\mathrm{sd}}$. The extreme case when a pair of subgap peaks reaches all the way to $V_{\mathrm{sd}}=0$ at two singlet-doublet parity crossings is shown in Fig.~\ref{fig14}b, and its effect on $I-V_{\mathrm{sd}}$ traces is shown in Fig.~\ref{fig14}a. Blue arrows indicate the subgap peak of largest conductance among the ones at positive and negative bias for a given gate voltage. To the left of state $D_{\mathrm{L}}$, this peak is at negative bias, whereas to the right it is at positive bias; i.e., the peak of largest conductance is antisymmetric with respect to the center of the state $D_{\mathrm{L}}$. Unsurprisingly, the same antisymmetry is observed for $I_\mathrm{0}$, with the negative-$I$ (positive-$I$) cusp occurring at larger $\abs{I}$ than its positive-$I$ (negative-$I$) counterpart to the left (right) of state $D_{\mathrm{L}}$. This occurs between the pairs of gate points marked by blue pins. This behavior contrasts with the hysteresis of underdamped Josephson junctions, where a sweep of $V_{\mathrm{sd}}$ from negative to positive values, as it is always the case in our device, would result in larger switching current for positive than negative $I$. Along the gate voltage delimited by the pair of red pins, the subgap states remain close to $V_{\mathrm{sd}}=0$, affecting the positive-negative symmetry of the $I$ cusps, which are just barely well-defined in this region. To the left and right of the parity crossings, in the $S$ states, YSR subgap peaks are far enough away from the supercurrent zero-bias conductance peak, and $I-V_{\mathrm{sd}}$ traces are consequently fully symmetric in $V_{\mathrm{sd}}$.

Figs.~\ref{fig14}c,d, show a different example with subgap peaks appearing near $V_{\mathrm{sd}}=0$ in Fig.~\ref{fig14}d. Nevertheless, these are far enough away to not affect the cusp symmetry in the respective $I-V_{\mathrm{sd}}$ traces in Fig.~\ref{fig14}c.

It is natural to expect that the appearance of such additional sub-gap conductance channels close to zero bias affects the resistance of the junction, and therefore the effective relation between $\langle I\rangle$ and $\langle V_{\mathrm{sd}}\rangle$. The modelling of this additional complication, however, is beyond the scope of this work.

\section*{Appendix D: Two-impurity model}

We model the double quantum dot system using the following superconducting generalization of the two-impurity Anderson impurity model:
\begin{equation}
H=\sum_{i} H_{\mathrm{imp},i} + H_\mathrm{LR} + \sum_{i} H_{\mathrm{BCS},i}
+\sum_{i} H_{\mathrm{hyb},i},
\label{FullHam}
\end{equation}
where the impurity Hamiltonians for quantum dots $i=L$ (left) and $i=R$ (right) with on-site energies $\epsilon_i$ and electron-electron repulsion (Hubbard) parameters $U_i$ are
\begin{equation}
H_{\mathrm{imp},i} = \frac{U_i}{2} \left( \hat{n}_i - n_i \right)^2,
\end{equation}
where the impurity occupancy (charge) operators are $\hat{n}_{i,\sigma}=d^\dag_{i,\sigma}
d_{i,\sigma}$, $\hat{n}_i = \hat{n}_{i,\uparrow}+\hat{n}_{i,\downarrow}$, while $n_i = 1/2 + \epsilon_i/U_i$ is the impurity energy level $\epsilon_i$ (controlled by the corresponding gate voltage) expressed in the dimensionless units of electron number. The inter-dot tunnelling amplitude, $t_\mathrm{d}$, enters through a tunnelling term,
\begin{equation}
H_{LR} = \sum_\sigma t_\mathrm{d} d^\dag_{L,\sigma} d_{R,\sigma} + \text{H.c.},
\end{equation}
the leads are modelled using BCS Hamiltonians
\begin{equation}
H_{\mathrm{BCS},i} = \sum_{k,\sigma} \epsilon_k c^\dag_{i,k\sigma} c_{i,k\sigma}
+ \sum_k \Delta c^\dag_{i,k\uparrow} c^\dag_{i,k\downarrow} + \text{H.c.},
\end{equation}
$U_\mathrm{d}$ is set to zero, and the QD-lead tunneling amplitudes, $t_\mathrm{{L/R}}$, enter through the hybridisation terms
\begin{equation}
H_{\mathrm{hyb},i} = t_i \sum_{k\sigma} c^\dag_{i,k\sigma} d_{i,\sigma} + \text{H.c.}
\end{equation}
In these expressions, $d_{i\sigma}$ are QD operators, while $c_{i,k\sigma}$ are lead operators. Finally, the tunnelling rates are defined as $\Gamma_i=\pi \rho_i t_i^2$, where $\rho_i$ is the density of states in lead $i$ in the normal state (i.e., in the absence of superconductivity).


\section*{Appendix E: 
Finite-bias conductance at weak inter-dot coupling}

For weak $t_\mathrm{d}$, $\mathrm{d}I/\mathrm{d}V_{sd}$ can be estimated
using the tunneling (Bardeen) approach, similar to that used in the analysis of scanning tunneling microscopy (STM) experiments. We assume that the voltage drops entirely across the weak link between the two dots, and that each quantum dot is in thermodynamic equilibrium with the neighboring lead. The chemical potential shifts in the leads are $\mu_\mathrm{{L,R}} = \pm eV_{\mathrm{sd}}/2$, and the local spectral functions on the dots are equally shifted. The superconductors are described in the semiconductor model, neglecting all coherence effects.
The current is then given by
\begin{equation}
\begin{split}
    I = \frac{4\pi e}{\hbar} |t_\mathrm{d}|^2 \int_{-\infty}^{\infty} [f(\omega-\mu_\mathrm{L})-f(\omega-\mu_\mathrm{R})] \\
    A_\mathrm{L}(\omega-\mu_\mathrm{L}) A_\mathrm{R}(\omega-\mu_\mathrm{R}) d\omega,
\end{split}
\end{equation}
where $f(x)=[1+\exp(x/k_BT)]^{-1}$ is the Fermi function. At zero temperature, this becomes
\begin{equation}
    I = \frac{4\pi e}{\hbar} |t_\mathrm{d}|^2 \int_{-eV_{\mathrm{sd}}/2}^{+eV_{\mathrm{sd}}/2}
     A_\mathrm{L}(\omega-eV_{\mathrm{sd}}/2) A_\mathrm{R}(\omega+eV_{\mathrm{sd}}/2) d\omega.
\end{equation}
We computed the spectral functions $A_{\mathrm{L,R}}(\omega)$ for the two subsystems with $t_\mathrm{d}=0$ using the density-matrix numerical renormalization group algorithm with  discretization parameter $\Lambda=2$, and broadened raw spectra using the broadening parameter $\alpha_\mathrm{b}=0.2$. After convolution of the spectral functions to obtain the current, $dI/dV_{\mathrm{sd}}$ was computed by taking numerical differences. This single-particle tunneling approach, valid only to second order order in $t_\mathrm{d}$, neglects Andreev reflections altogether and presumes that relaxation rates of quasiparticles in the sub-gap states are larger than the inter-dot tunneling rate.


\begin{figure} [h!]
\includegraphics[width=1\linewidth]{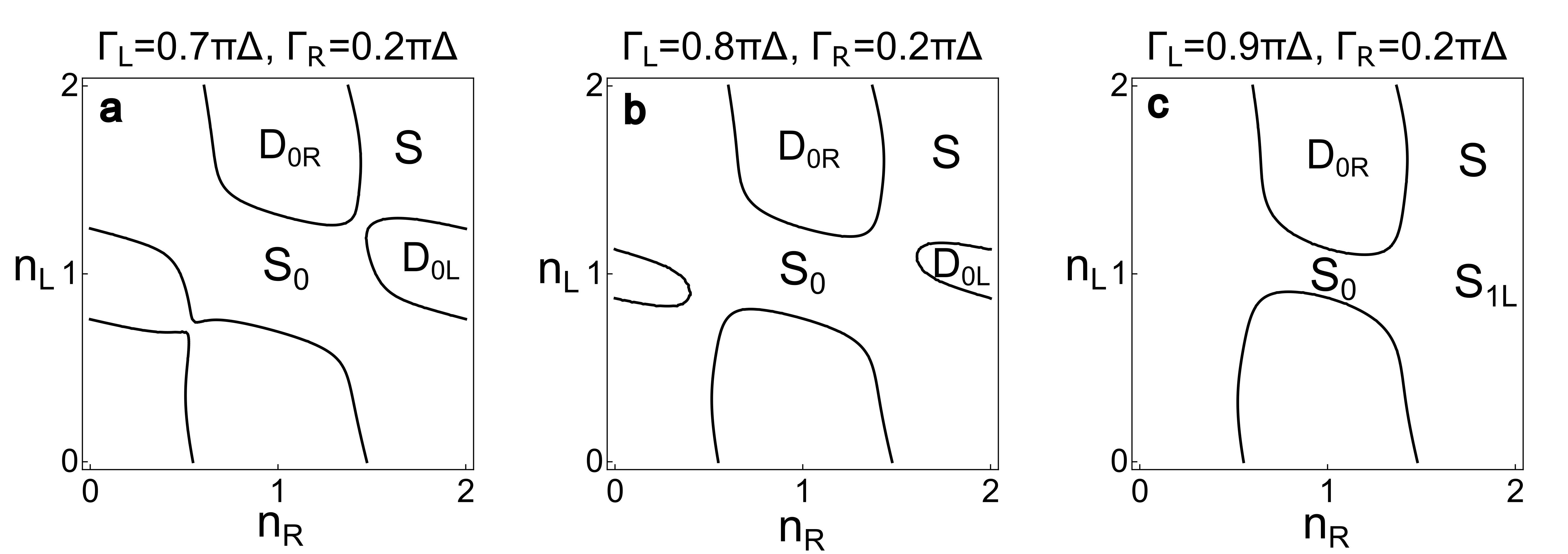}
\caption{
Parity transition lines calculated using a ZBW approximation with a gate-dependent tunnel coupling $t_\mathrm{R}$. Black lines indicate changes of GS parity. Increasing $\Gamma_\mathrm{L}$ from (a) to (c). These figures should be qualitatively compared with Fig.~\ref{fig6} of the main text.
}
\label{Tfig1}
\end{figure}

\begin{figure} [h!]
\includegraphics[width=1\linewidth]{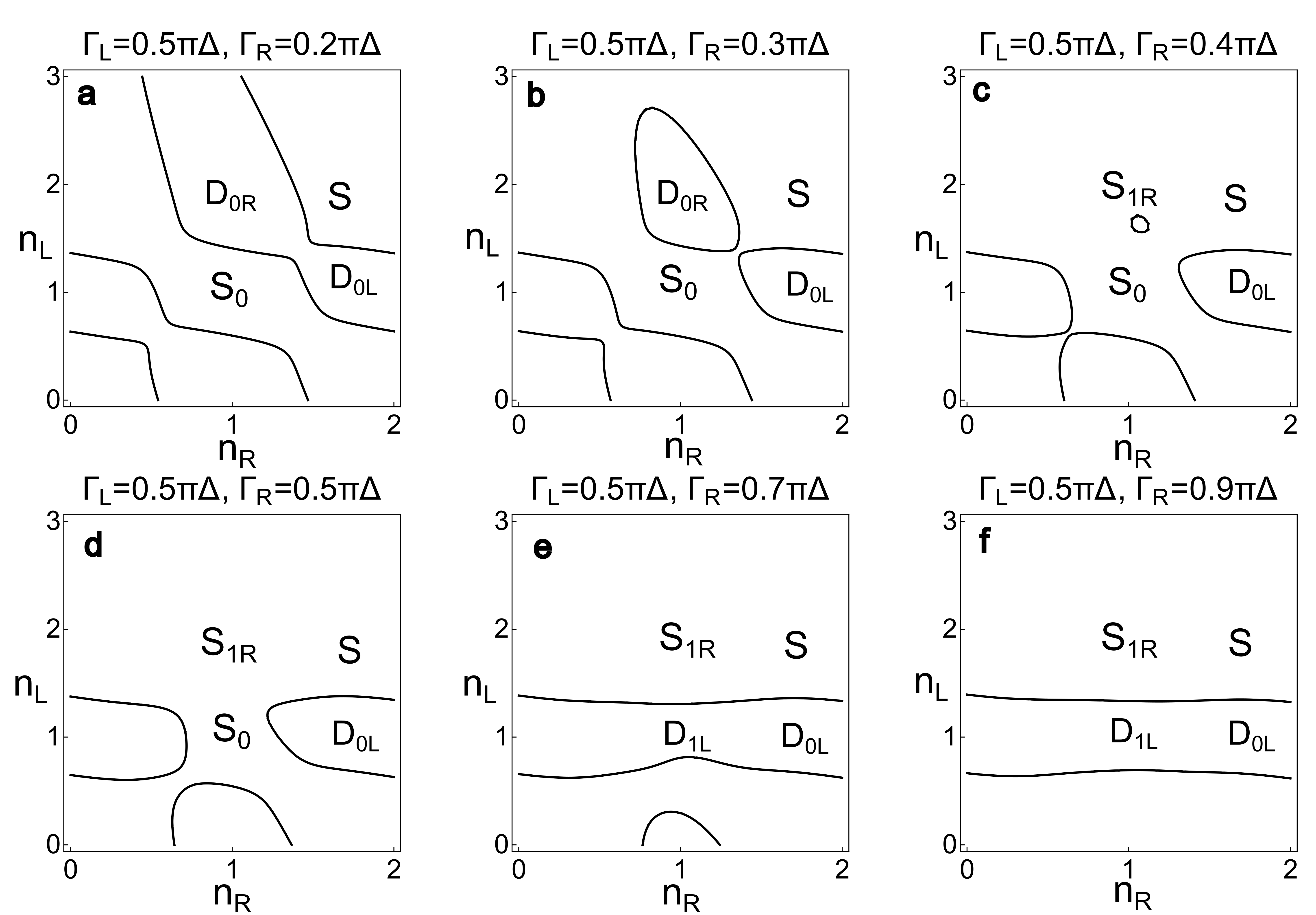}
\caption{
Parity transition lines calculated using a ZBW approximation with a gate-dependent tunnel coupling $t_\mathrm{R}$. Black lines indicate changes of GS parity.  Increasing $\Gamma_\mathrm{R}$ from (a) to (f). These figures should be qualitatively compared with Fig.~\ref{fig7} of the main text.
}
\label{Tfig2}
\end{figure}

\section*{Appendix F: Effects of finite inter-dot charging energy and gate-dependent tunnel coupling}

In this section we investigate the effects of a possible influence of $V_{\mathrm{gL}}$ on $t_\mathrm{R}$. For efficient parameter exploration, we do this using a zero-bandwidth (ZBW) approximation~\cite{grove2017yu} of the superconducting leads in Eq.~(\ref{FullHam}), taking $H_{\mathrm{BCS_i}}\approx \sum_i \Delta c^\dagger_{i\uparrow}c^\dagger_{i\downarrow}+\operatorname{h.c.}$ and numerically diagonalizing the Hamiltonian. We model the tunnel couplings, $t_i$, as $t_\mathrm{L} = \sqrt{2\Delta\Gamma_\mathrm{L}/2}$ and $t_\mathrm{R} = \sqrt{2\Delta\Gamma_\mathrm{R}/2}(1+n_\mathrm{L}/4)$, where $t_\mathrm{R}$ now depends on $V_{\mathrm{gL}}$ through $n_\mathrm{L}$. To more realistically mimic the experimental scenario we also include an inter-dot Coulomb repulsion, $U_{\mathrm{d}}$, such that $H_\mathrm{imp}=\sum_{i=\mathrm{L,R}}U_i/2(\hat{n}_{i}-n_i)^2+U_{\mathrm{d}}(\hat{n}_\mathrm{L}-n_\mathrm{L})(\hat{n}_\mathrm{R}-n_\mathrm{R})$ with $U_{\mathrm{d}}=0.5$ meV and $\hat{n}_i=\hat{n}_{i\uparrow}+\hat{n}_{i\downarrow}$ being the electron counting operator. Other parameters are identical to parameters used for the NRG calculations presented above. Comparing Fig.~\ref{Tfig2}a to Fig.~\ref{fig5}a, the main effect of $U_\mathrm{d}$ is to provide an angle on the $n_\mathrm{R}$, $n_\mathrm{L}$ parity lines.

\begin{figure} [t!]
\includegraphics[width=1\linewidth]{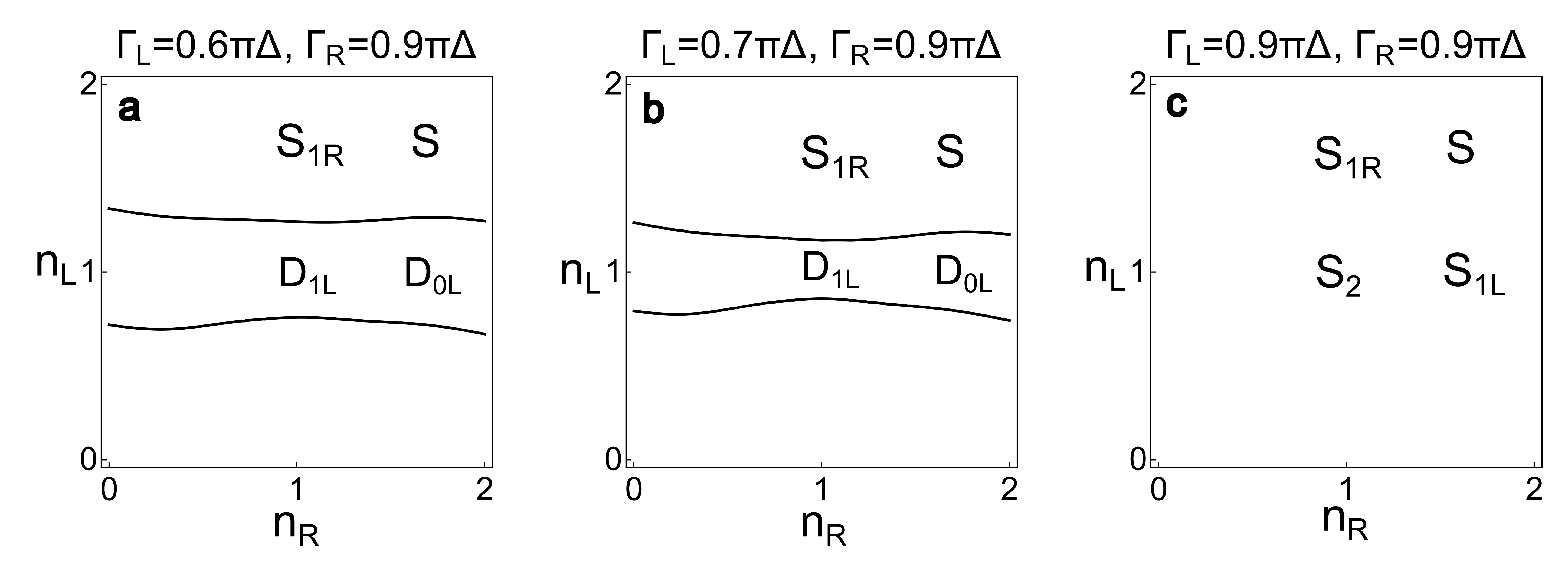}
\caption{
Parity transition lines calculated using a ZBW approximation with a gate-dependent tunnel coupling $t_\mathrm{R}$. Black lines indicate changes of GS parity. Increasing $\Gamma_\mathrm{L}$ from (a) to (c). These figures should be qualitatively compared with Fig.~\ref{fig8} of the main text.
}
\label{Tfig3}
\end{figure}

In Figs.~\ref{Tfig1}, \ref{Tfig2}, and \ref{Tfig3} we show GS parity transition lines (black lines) of the S-DQD-S charge stability diagrams for various values of $\Gamma_\mathrm{L}$ and $\Gamma_\mathrm{R}$, corresponding qualitatively to data in Figs.~\ref{fig6}, \ref{fig7} and \ref{fig8}. The gate-dependent coupling is seen to account in Fig.~\ref{Tfig2} for the asymmetric screening around the electron-hole symmetry point occurring when $\Gamma_\mathrm{L}$ is increased, which is necessary to explain the experimental findings in Fig.~\ref{fig7}, that is, the observation that the doublet to singlet transition reflected by the vanishing of the two parity transition lines happens faster on the top ($V_{\mathrm{gL}}\approx -2.1$V) than at the bottom ($V_{\mathrm{gL}}\approx -2.3$V).


\end{document}